\documentclass[%
nofootinbib,
amsmath,amssymb,
]{aastex631}

\usepackage{graphicx}
\usepackage{dcolumn}
\usepackage{bm}
\usepackage{braket}
\usepackage{hyperref}
\usepackage{xcolor, soul}
\usepackage{pygtex}
\usepackage{physics, amsmath}
\usepackage{inputenc}

\usepackage[acronym]{glossaries}
\makeglossaries
\newacronym[plural=GWs,firstplural=gravitational waves (GWs)]{gw}{GW}{gravitational wave}
\newacronym[plural=BHs,firstplural=black holes (BHs)]{bh}{BH}{black hole}
\newacronym[plural=CBCs,firstplural=compact binary coalescences (CBCs)]{cbc}{CBC}{compact binary coalescence}
\newacronym[plural=SGWBs,firstplural=stochastic gravitational-wave backgrounds (SGWBs)]{sgwb}{SGWB}{stochastic gravitational-wave background}
\newacronym[plural=GWBs, firstplural=gravitational-wave backgrounds (GWBs)]{gwb}{GWB}{gravitational-wave background}
\newacronym[plural=DFTs, firstplural=discrete Fourier transforms (DFTs)]{dft}{DFT}{discrete Fourier transform}
\newacronym[plural=MVUEs, firstplural=minimum-variance unbiased estimators(MVUEss)]{mvue}{MVUE}{minimum-variance unbiased estimator}
\newacronym{ligo}{LIGO}{Laser Interferometer Gravitational-wave Observatory}
\newacronym{lvk}{LVK}{LIGO, Virgo, and KAGRA Collaboration}
\newacronym{lisa}{LISA}{Laser Interferometer Space Antenna}
\newacronym{et}{ET}{Einstein Telescope}
\newacronym{ce}{CE}{Cosmic Explorer}
\newacronym[plural=BBHs,firstplural=binary black holes (BBHs)]{bbh}{BBH}{binary black hole}
\newacronym[plural=BNSs,firstplural=binary neutron stars (BNSs)]{bns}{BNS}{binary neutron star}
\newacronym[plural=MDSCs,firstplural=mock data and science challenges (MDSCs)]{mdsc}{MDSC}{mock data and science challenges}
\newacronym[plural=IFTed,firstplural=inverse--discrete Fourier transformed (IFTed)]{ift}{IFT}{inverse--discrete Fourier transform}
\newacronym{plpp}{PLPP}{power-law-plus-peak}
\newacronym{pl}{PL}{power-law}
\newacronym{snr}{SNR}{signal-to-noise ratio}
\newacronym{psd}{PSD}{power spectral density}
\newacronym[plural=ORFs, firstplural=overlap reduction functions (ORFs)]{orf}{ORF}{overlap reduction function}
\newacronym{asd}{ASD}{amplitude spectral density}
\newacronym{csd}{CSD}{cross-spectral density}
\newacronym{gwosc}{GWOSC}{Gravitational-wave Open Science Center}
\newacronym{ks}{KS}{Kolmogorov-Smirnov}
\newacronym{gr}{GR}{General Relativity}
\newacronym{pi}{PI}{power-law integrated sensitivity}

\begin{document}
\nolinenumbers
\title{{\tt pygwb}: Python-based library for gravitational-wave background searches}

\author{Arianna I. Renzini}
\email{arenzini@caltech.edu}
\affiliation{LIGO Laboratory,  California  Institute  of  Technology,  Pasadena,  California  91125,  USA}
\affiliation{Department of Physics, California Institute of Technology, Pasadena, California 91125, USA}
\author{Alba Romero-Rodríguez}
 \affiliation{Theoretische Natuurkunde, Vrije Universiteit Brussel, Pleinlaan 2, B-1050 Brussels, Belgium} 
\author{Colm Talbot}
\affiliation{Kavli Institute for Astrophysics and Space Research, Massachusetts Institute of Technology, 77 Massachusetts Ave, Cambridge, MA 02139, USA}
\author{Max Lalleman}
 \affiliation{Universiteit Antwerpen, Prinsstraat 13, 2000 Antwerpen, Belgium} 
\author{Shivaraj Kandhasamy}
\affiliation{Inter-University Centre for Astronomy and Astrophysics, Pune 411007, India}
\author{Kevin Turbang}
\affiliation{Universiteit Antwerpen, Prinsstraat 13, 2000 Antwerpen, Belgium}
\affiliation{Theoretische Natuurkunde, Vrije Universiteit Brussel, Pleinlaan 2, B-1050 Brussels, Belgium} 
\author{Sylvia Biscoveanu}
\affiliation{Kavli Institute for Astrophysics and Space Research, Massachusetts Institute of Technology, 77 Massachusetts Ave, Cambridge, MA 02139, USA}
\affiliation{LIGO Laboratory, Massachusetts Institute of Technology, 185 Albany St, Cambridge, MA 02139, USA}
\author{Katarina Martinovic}
\affiliation{Theoretical Particle Physics and Cosmology Group,  Physics Department, \\ King's College London, University of London, Strand, London WC2R 2LS, United Kingdom}
\author{Patrick Meyers}
\affiliation{Theoretical Astrophysics Group, California Institute of Technology, Pasadena, CA 91125, USA}
\author{Leo Tsukada}
\affiliation{Department of Physics, The Pennsylvania State University, University Park, Pennsylvania 16802, USA}
\affiliation{Institute for Gravitation and the Cosmos, The Pennsylvania State University, University Park, Pennsylvania 16802, USA}
\author{Kamiel Janssens}
 \affiliation{Universiteit Antwerpen, Prinsstraat 13, 2000 Antwerpen, Belgium}
 \affiliation{Universit\'e C$\hat{o}$te d’Azur, Observatoire C$\hat{o}$te d’Azur, ARTEMIS, Nice, France}
\author{Derek Davis}
\affiliation{LIGO Laboratory,  California  Institute  of  Technology,  Pasadena,  California  91125,  USA}
\affiliation{Department of Physics, California Institute of Technology, Pasadena, California 91125, USA}
\author{Andrew Matas}
\affiliation{Max Planck Institute for Gravitational Physics (Albert Einstein Institute), D-14476 Potsdam, Germany}
\author{Philip Charlton}
\affiliation{OzGrav, Charles Sturt University, Wagga Wagga, New South Wales 2678, Australia}
\author{Guo-Chin Liu}
\affiliation{Department of Physics, Tamkang University, Danshui Dist., New Taipei City 25137, Taiwan}
\author{Irina Dvorkin}
\affiliation{Institut d’Astrophysique de Paris, Sorbonne Universit\'e \& CNRS, UMR 7095, 98 bis bd Arago, F-75014 Paris, France}
\affiliation{Université Paris Cité, CNRS, Astroparticule et Cosmologie, F-75013 Paris, France}
\author{Sharan Banagiri}
\affiliation{Center for Interdisciplinary Exploration and Research in Astrophysics (CIERA), Northwestern University, 1800 Sherman Ave, Evanston, IL 60201, USA}
\author{Sukanta Bose}
\affiliation{Inter-University Centre for Astronomy and Astrophysics, Pune 411007, India}
\author{Thomas Callister}
\affiliation{Kavli Institute for Cosmological Physics, The University of Chicago, 5640 S. Ellis Ave., Chicago, IL 60615, USA}
\author{Federico De Lillo}
\affiliation{Centre for Cosmology, Particle Physics and Phenomenology (CP3),\\
Universit\'e catholique de Louvain, Louvain-la-Neuve, B-1348, Belgium}
\author{Luca D'Onofrio}
\affiliation{Universit\`a di Napoli "Federico II", Dipartimento di Fisica "Ettore Pancini", Compl. Univ. di Monte S. Angelo, Via Cinthia 21, I-80126, Napoli, Italy}
\affiliation{INFN, Sezione di Napoli, Compl. Univ. di Monte S. Angelo, Edificio G, Via Cinthia, I-80126, Napoli, Italy}
\author{Fabio Garufi}
\affiliation{Universit\`a di Napoli "Federico II", Dipartimento di Fisica "Ettore Pancini", Compl. Univ. di Monte S. Angelo, Via Cinthia 21, I-80126, Napoli, Italy}
\affiliation{INFN, Sezione di Napoli, Compl. Univ. di Monte S. Angelo, Edificio G, Via Cinthia, I-80126, Napoli, Italy}
\author{Gregg Harry}
\affiliation{Physics Department, American University, Washington, DC 20016, USA}
\author{Jessica Lawrence}
\affiliation{Department of Physics, Texas Tech University, Lubbock, TX 79409, USA}
\author{Vuk Mandic}
\affiliation{School of Physics and Astronomy, University of Minnesota, Minneapolis, MN 55455, USA}
\author{Adrian Macquet}
\affiliation{Institut de F\'{\i}sica d’Altes Energies (IFAE), The Barcelona Institute of Science and Technology, Campus UAB, 08193 Bellaterra (Barcelona) Spain}
\author{Ioannis Michaloliakos}
\affiliation{Department of Physics, University of Florida, Gainesville, Florida 32611, USA}
\author{Sanjit Mitra}
\affiliation{Inter-University Centre for Astronomy and Astrophysics, Pune 411007, India}
\author{Kiet Pham}
\affiliation{School of Physics and Astronomy, University of Minnesota, Minneapolis, MN 55455, USA}
\author{Rosa Poggiani}
\affiliation{Universit\`a di Pisa, I-56127 Pisa, Italy}
\affiliation{INFN, Sezione di Pisa, I-56127 Pisa, Italy}
\author{Tania Regimbau}
\affiliation{LAPP, CNRS, 9 Chemin de Bellevue, 74941 Annecy-le-Vieux, France}
\author{Joseph D. Romano}
\affiliation{Department of Physics, Texas Tech University, Lubbock, TX 79409, USA}
\author{Nick van Remortel}
 \affiliation{Universiteit Antwerpen, Prinsstraat 13, 2000 Antwerpen, Belgium}
\author{Haowen Zhong}
\affiliation{School of Physics and Astronomy, University of Minnesota, Minneapolis, MN 55455, USA}

\begin{abstract}
The collection of gravitational waves (GWs) that are either too weak or too numerous to be individually resolved is commonly referred to as the gravitational-wave background (GWB). 
A confident detection and model-driven characterization of such a signal will provide invaluable information about the evolution of the Universe and the population of GW sources within it. %
We present a new, user-friendly Python--based package for gravitational-wave data analysis to search for an isotropic GWB in ground--based interferometer data. %
We employ cross-correlation spectra of GW detector pairs to construct an optimal estimator of the Gaussian and isotropic GWB, and Bayesian parameter estimation to constrain GWB models. %
The modularity and clarity of the code allow for both a shallow learning curve and flexibility in adjusting the analysis to one's own needs. %
We describe the individual modules which make up {\tt pygwb}, following the traditional steps of stochastic analyses carried out within the LIGO, Virgo, and KAGRA Collaboration. %
We then describe the built-in pipeline which combines the different modules and validate it with both mock data and real GW data from the O3 Advanced LIGO and Virgo observing run. %
We successfully recover all mock data injections and reproduce published results. %

\end{abstract}



\section{Introduction}

Since the first direct \gls{gw} detection~\cite{PhysRevLett.116.061102}, the field of \gls{gw} astrophysics has exploded, now encompassing a wide range of instrumental and observational campaigns across the globe. %
These detection efforts monitor a vast range of frequencies, from the nanohertz to the kilohertz, and are sensitive to a multitude of \gls{gw} sources emitting therein. %
While the \gls{gw} sources in each band may present extremely different characteristics, a potential candidate for {\it all} \gls{gw} measurements is a \gls{gwb}, given by the collection of all \glspl{gw} too faint to be individually resolved, or by the incoherent overlap of a large number of signals in the same band~\cite{Regimbau:2011rp, Christensen_review, SGWB_review2022_AIR}. %
This sort of signal has been targeted in several different datasets~\cite{LIGO_S1, LIGO_S4, LIGO_S5, LIGO_O1, LIGO_O2, LIGO_O3} using search methods which estimate the \gls{gw} strain signal power modelling the signal as stochastic, frequently resorting to cross-correlation of multiple independent observations~\cite{Allen:1999}. %
These searches are often referred to as stochastic searches by the \gls{gw} detection community, and these backgrounds are often referred to as \glspl{sgwb}, even though, in practice, not all target background signals are fully described by stochastic variables\footnote{To avoid confusion, in this paper we will use the term \gls{sgwb} to refer to signals that are indeed defined as stochastic fields.}, and this definition may imply an approximation. %
So far, no confident detection of a \gls{gwb} has been claimed.

With this paper we present {\tt pygwb}~\cite{pygwb_pypi}, a new Python--based package tailored to searches for isotropic \glspl{gwb} with current ground-based interferometers, namely the \gls{ligo}~\cite{aLIGO_sensitivity}, the Virgo observatory~\cite{Acernese_2014}, and the KAGRA detector~\cite{KAGRA_design_doc}, and with the potential to be expanded and adapted to several other detection efforts. %
The core analysis tools, described in detail in what follows, are heavily inspired by the \gls{lvk} stochastic analysis code, {\tt stochastic.m}. %
The latter consists of a set of {\tt MATLAB} scripts easily parallelizable on a high-throughput computing cluster, and has been used in \gls{lvk} data analysis for the past data acquisition runs~\cite{LIGO_S4, LIGO_S5, LIGO_O1, LIGO_O2, LIGO_O3}. %
These include the 3 observing runs: O1 (September 2015 to January 2016), O2 (November 2016 to August 2017), and O3 (April 2019 to March 2020), performed with Advanced \gls{ligo} Hanford and Livingston, and Advanced Virgo~\cite{Acernese_2014} for part of O2 and O3. %
Data from Virgo has been included in stochastic analyses as of the latest observing run. %
The analysis consists in the calculation of an optimal statistic~\cite{Allen:1999} from the data of multiple interferometers, which is directly related to the amplitude of the \gls{gwb} signal.

A notable change throughout the years of stochastic \gls{gw} analyses has been the constant shift towards Bayesian parameter estimation~\cite{Mandic_2012, LIGO_O3}. %
To date, there is no preferred stochastic parameter estimation software, and different groups have employed private scripts. %
To extend the scope of the stochastic search beyond the optimal statistic, we include a parameter estimation module in {\tt pygwb} based on the {\tt Bilby} package~\cite{Ashton:2018jfp} which allows the user to test both predefined and user-defined models and obtain posterior distributions on the parameters of interest. %

The steady inflow of ever-improving \gls{gw} data open for analysis~\cite{gwosc} has been a catalyst for open-source \gls{gw} data analysis codebase development. %
By adopting the Python language and focusing on user-friendliness, flexibility, and portability, we intend to introduce stochastic searches to the wider \gls{gw} community. %
Detecting a \gls{gwb} with ground-based interferometers will be a community effort, and we expect search pipelines to evolve along the way. %
The format and structure of {\tt pygwb} facilitates this evolution, and conversely, the package is suitable for beginners approaching \gls{gwb} data analysis for the first time. %


This paper is structured as follows. %
In Sec. \ref{sec: GWB analysis}, concepts related to the characterization and detection methods of a \gls{gwb} are reviewed. %
A detailed overview of the individual modules that make up the {\tt pygwb} package follows in Sec. \ref{sec: modules}, outlining the steps of \gls{lvk} stochastic analyses. %
Several manager objects which store relevant data and handle the analysis internally are described in Sec.~\ref{sec: manager objects}. %
The built-in {\tt pygwb} pipeline which combines individual modules and performs the search for an isotropic \gls{gwb} is presented in Sec.~\ref{sec: pipeline}. %
To test the capabilities of the pipeline, mock datasets with a variety of simulated signals are analyzed in Sec.~\ref{sec: MDC}. %
To conclude, results from the analysis of the third \gls{lvk} collaboration observing run, O3, are presented and compared with collaboration results in Sec.~\ref{Sec:O3Data}.
\section{The isotropic stochastic analysis}
\label{sec: GWB analysis}

A \gls{sgwb} is characterized by its spectral emission, which is the target of stochastic \gls{gw} searches. %
    The spectrum is typically parametrized by the \gls{gw} fractional energy density spectrum $\Omega_{\rm GW}(f)$, such that
\begin{equation}
	\Omega_{\rm GW}(f) = \frac{1}{\rho_\mathrm{c}} \frac{\text{d}\rho_{\rm GW} (f)}{\text{d}\ln f}\,,
\label{eq:Omegarho}
\end{equation}
where $\text{d}\rho_{\rm GW}$ is the energy density of \glspl{gw} in the frequency band $f$ to $f+\text{d}f$, and $\rho_{\rm c}$ is the critical energy density in the Universe. When integrated over $\text{d}\log f$, $\Omega_{\rm GW}(f)$ gives the total dimensionless \gls{gw} energy density. %
The $\Omega_{\rm GW}(f)$ spectrum is thus directly related to the intensity of \glspl{gw}. %
Specifically, from Eq.~\eqref{eq:Omegarho} it follows that~\cite{Allen:1999}
\begin{equation}
	\Omega_{\rm GW}(f) = \frac{4\pi^2f^3}{\rho_\mathrm{c} G}S_h(f)\,,
\label{eq:omegatoI}
\end{equation}
where the strain spectral density $S_h(f)$ is defined as the polarization--averaged second moment of the stochastic \gls{gw} strain field, decomposed into its polarization components $h_{+}$ and $h_{\times}$,
\begin{equation}
\langle h^{}_+(f,\,\hat{\bm n})\,h^\ast_+(f',\,\hat{\bm n}')\rangle + \langle h^{}_\times(f,\,\hat{\bm n})\,h^\ast_\times(f',\,\hat{\bm n}')\rangle  = \,\delta^{(2)}(\bm{n},\bm{n'})\,\delta(f-f')\,
 S_h(f,\,\hat{\bm n})\,,
\label{eq:intensity}
\end{equation}
assuming statistical homogeneity. %
The unit vectors $\hat{\bm n}$, $\hat{\bm n}'$ span the 2-sphere, while $f\in \mathbb{R}$. %
In the plane wave formalism, $h_{+}$ and $h_{\times}$ in Eq.~\eqref{eq:intensity} are the Fourier coefficients of the time-domain strain fields. %
If these are stochastically distributed, these give rise to a \gls{sgwb} which we describe solely through the statistical moments of the distribution. %
In particular, a Gaussian \gls{sgwb} is fully described by its second moments, hence the spectral density in Eq.~\eqref{eq:intensity} is the primary target of a search which assumes the signal to be both stochastic and Gaussian. %
More details on these quantities can be found for example in~\cite{Romano_2017}. %

Laser interferometers such as \gls{ligo} and Virgo are sensitive to the strain field in the time domain coming from all directions, $h(t)$. %
These detectors measure the \gls{gw} strain filtered through a linear response function $F$ (see definition in~\cite{Romano_2017}) plus a detector noise component $n$, which we may write in shorthand as
\begin{equation}
    d(t) = F(t) \star h(t) + n(t),
\end{equation}
where ``$\star$'' indicates a convolution operation. %
Given that the \gls{sgwb} signal is weak and hard to distinguish from instrumental noise, cross-correlating two independent, time-coincident datastreams with uncorrelated noise is an effective way to construct an estimator for $\Omega_{\rm GW}(f)$. %
We assume our target stochastic \gls{gw} signal is stationary, Gaussian, and isotropic. %
We further assume the detector noise is Gaussian and uncorrelated between detectors, which is a fair assumption in the case of ground-based interferometers at current detector sensitivity\footnote{In future detectors, correlated noise will become a significant problem, and quite a few methods for mitigating it have been proposed, including Wiener filtering and Bayesian parameter estimation~\cite{ThraneChristensen2013,ThraneChristensen2014,CoughlinChristensen2016,HimemotoTaruya2017,CoughlinCirone2018,HimemotoTaruya2019,Meyers_2020,JanssensMartinovic2021,JanssensBall2023,HimemotoNishizawa2023}.} (after specific mitigation)~\cite{LIGO_O3,JanssensBall2023}, and that the noise amplitude is much larger than the signal amplitude. %
Under these assumptions\footnote{Failure of stationarity or Gaussianity implies the estimator is sub-optimal, yet still valid~\cite{Drasco-Flanagan:2003, Lawrence:2023buo}; failure of isotropy would also induce a bias, and the target signal would be ill-defined~\cite{PhysRevD.107.023024}.}, it has been shown~\cite{SGWBH1H2, Romano_2017} that the cross-correlation--based \gls{mvue} of $\Omega_{\rm GW}$ at a frequency bin $f$ and the corresponding variance is given by,
\begin{equation}\label{eq:Omega}
    \hat{\Omega}_{{\rm GW}, f} = \frac{\Re[C_{IJ, f}]}{\gamma_{IJ}(f) S_0(f)} \ ,
\end{equation}
and
\begin{equation}\label{eq:Variance}
    \sigma^2_{{\rm GW,} f} = \frac{1}{2 T \Delta f} \frac{P_{I, f} P_{J, f}}{\gamma^2_{IJ}(f) S^2_0(f)},
\end{equation}
where $C_{IJ, f}$ is the one-sided \gls{csd} and $P_{I, f}$ is the one-sided (auto-)\gls{psd} of strain data $d_t$ from two detectors $(I,J)$, as defined below in Sec.~\ref{sec:spectral}\footnote{Note that, in previous works, the notation $C_{IJ}$ was used to define the cross-correlation statistic itself \cite{LIGO_O3}. This is not the case in this paper.}. %
Note that throughout this work we will denote continuous functions of the frequency with the notation $(f)$, whereas discrete functions of the frequency will be denoted with a subscript $_f$. %
Typically, in this paper, discrete functions of frequency are estimators for continuous functions, and in Equations such as Eqs.~\eqref{eq:Omega} and ~\eqref{eq:Variance} which mix discrete and continuous functions our notation implies that continuous functions are evaluated at the discrete set of frequencies for which we know the value of the discrete functions. %
In the above, $T$ is the duration of data used to produce the above spectral densities, and $\gamma_{IJ}(f)$ is the cross-correlated \gls{gw} response, or \gls{orf}, which is the polarization-- and sky-- averaged cross-correlation of the individual detector responses, $F_I$. The \gls{orf} normalized for a pair of perpendicular-arm interferometers is given by~\cite{Allen:1999}
\begin{equation}
    \label{eq:orf}
    \gamma_{IJ}(f)  = \frac{5}{8\pi} \sum_{A}\int_{S^2} d \hat{\bm n} F^A_I(f, \hat{\bm n})F^A_J(f, \hat{\bm n}) e^{-i 2\pi f \hat{\bm n}\cdot({\bm x}_I-{\bm x}_J)}\,,
\end{equation}
where $\hat{\bm n}$ is the unit vector on the sky, in an arbitrary basis\footnote{The \gls{orf} in {\tt pygwb} is calculated in geocentric coordinates.}, ${\bm x}_I-{\bm x}_J$ is the difference between the position vectors of the two detectors $I$ and $J$ respectively, and $A$ spans the polarization basis. %
The \gls{orf} quantifies the reduction in sensitivity of the cross-correlation stochastic search due to the detectors not being co-aligned and co-located, and having different non-trivial responses. %
The function $S_0$ is defined as~\cite{Romano_2017, SGWB_review2022_AIR}
\begin{equation}
    S_0(f) = \frac{3H_0^2}{10\pi^2}\frac{1}{f^3},
    \label{eq:S0}
\end{equation}
and converts a \gls{gw} strain power spectrum into a fractional energy density. %
The derivation of $S_0$ is shown in~\cite{Allen:1999}, and note that it includes the normalization factor of the \gls{orf}, $5/8\pi$, which ensures $\gamma_{IJ}(f)\equiv 1$ for co-aligned, co-located detectors.

There are two important considerations to make regarding the estimator in Eqs.~\eqref{eq:Omega} and~\eqref{eq:Variance}. %
Firstly, the implementation of a \gls{dft} over a finite time $T$ in the estimator of the continuous non-periodic quantity $\Omega_{\rm GW}(f)$ may create spectral artifacts, as seen in~\cite{numerical_recipes2007, whelan_CC_dcc}. %
We outline how this is handled in Sec.~\ref{sec:spectral}. %
Secondly, as the estimator is initially derived as a minimal variance estimator in the time domain~\cite{Allen:1999}, the narrow-band frequency estimator in Eq.~\eqref{eq:Omega} is actually obtained from a broad-band one, as will be clarified in Sec.~\ref{sec:postproc}. %
In the rest of this paper, we refer to $\hat{\Omega}_{{\rm GW}, f}$ as the optimal estimator of the signal spectrum ${\Omega}_{\rm GW} (f)$. %
The optimality of the estimator can either be justified by the proof that this is an \gls{mvue}, or equivalently by showing that it maximizes a reasonable likelihood for the data. When performing parameter estimation as outlined in Sec.~\ref{sec:pe}, we in fact employ a Gaussian likelihood which is maximized by $\hat{\Omega}_{{\rm GW}, f}$.

In stochastic analyses with current interferometers, we take advantage of long observing times to improve detection statistics. %
In practice, the data are segmented into smaller chunks and analyzed individually before they are optimally combined to produce an estimate. This is convenient due to potential non-stationarities in the detector noise over both short time-scales, such as the length of an individual data segment, and long time-scales, such as the total observation time, as well as reducing computational costs. %
Assuming each time segment is independent, we perform a weighted average over all segments to calculate $\hat{\Omega}_{{\rm GW}, f}$ for long observations. %
This average can be thought of as an approximation to the ensemble averages in Eq.~\eqref{eq:intensity}. Hence the more independent observations are averaged over, the better the measurement. The averaging procedure is described in full in Sec.~\ref{sec:postproc}.

The narrow-band statistic of Eqs.~\eqref{eq:Omega} and~\eqref{eq:Variance} assumes each frequency bin is independent. The information from each bin can be combined under the assumption of a known \gls{gw} spectral density distribution. In \gls{gwb} analyses, it is most common to assume a power-law spectral shape for $\Omega_{\rm GW}$,
\begin{equation}
    \Omega_{\rm GW}(f) = \Omega_{\rm ref}\left(\frac{f}{f_{\rm ref}}\right)^{\alpha}\,,
    \label{eq:omega-plaw}
\end{equation}
where $\alpha$ is the spectral index of the signal, and $f_{\rm ref}$ is a reference frequency, and $\Omega_{\rm ref}$ is defined as $\Omega_{\rm ref}\equiv\Omega_{\rm GW}(f_{\rm ref})$. %
Under this assumption, the rescaling
\begin{equation}
    \label{eq:Salpha}
    H_{\rm ref, \alpha}(f) = \qty(\frac{f}{f_{\rm ref}})^\alpha
\end{equation}
can be used to re-weight the estimate of the spectrum $\hat\Omega_{{\rm GW}, f}$, obtained for $\alpha=0$, to optimize the statistic for a specific spectral index $\alpha$ at a chosen reference frequency $f_{\rm ref}$, reducing the search to the estimation of a single number, $\Omega_{\rm ref}$. This procedure is referred to as {\it re-weighting} and is clarified in Sec.~\ref{sec:postproc}. %
Alternatively, it is also possible to keep $\alpha$ as a free parameter in the analysis, and estimate both $\Omega_{\rm ref}$ and $\alpha$ from the data. This is described in Sec.~\ref{sec:pe}.

\section{Individual modules}
\label{sec: modules}
What follows is a detailed step-by-step presentation of the stochastic analysis pipeline. %
We follow the natural structure of the code for clarity as we introduce each module individually. %
To start, we present the {\tt preprocessing} module which pre-conditions the time-domain strain from \gls{gw} detectors for spectral analysis. %
In {\tt spectral}, we explain the power spectrum and cross-spectrum calculations, which produce the $P_{I, f}$ and $C_{IJ, f}$ spectra in Eqs.~\eqref{eq:Omega} and~\eqref{eq:Variance}. %
We then describe {\tt postprocessing}, which includes the averaging procedures employed over large datasets to obtain an optimal estimate of the signal amplitude starting from the quantities in Eqs.~\eqref{eq:Omega} and~\eqref{eq:Variance}, and knowledge of the expected spectral shape. %
In {\tt delta-sigma cut} and {\tt notch}, we present modules which focus on data quality checks, and the implementation of relevant time-domain and frequency-domain data cuts. %
We then describe the built-in parameter estimation module {\tt pe}, based on {\tt Bilby}~\cite{Ashton:2018jfp}, a Python--based Bayesian inference library widely used in GW data analysis. %
Finally, we present the {\tt simulator} module, which includes different mock-data injection techniques for \gls{gwb} study and detection validation. %

\begin{figure}
    \centering
    \includegraphics[width = 0.65\textwidth]{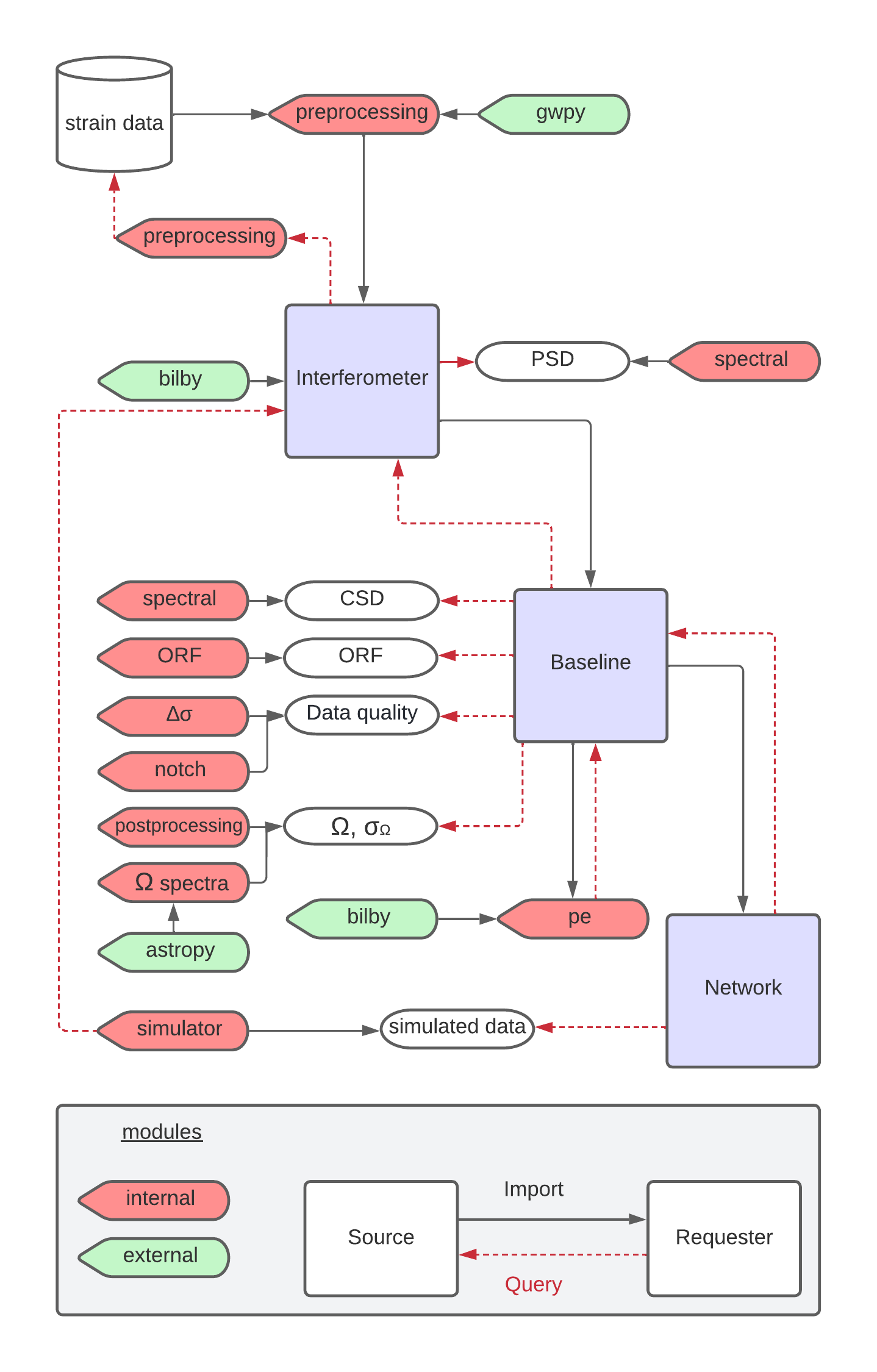}
    \caption{Schematic overview of {\tt pygwb} analysis flow. In blue squares, we show the manager objects of the code that handle the analysis internally. These manager objects query (red arrows) different modules for specific objects, calculations, or quantities (rounded bubbles), imported (grey arrows) by either internal (i.e., within {\tt pygwb}) or external modules (i.e., outside of {\tt pygwb}). Internal modules are indicated in red, while external modules are indicated in green. }
    \label{fig:flowchart}
\end{figure}

A schematic of the {\tt pygwb} package is presented in Fig.~\ref{fig:flowchart}. %
This includes the manager objects {\tt Interferometer}, {\tt Baseline}, and {\tt Network}, presented in Sec.~\ref{sec: manager objects}.

\subsection{\tt preprocessing}\label{sec:preproc}
Pre-processing is the first step of stochastic GW data analysis, in which data are read, downsampled and high-pass filtered. %
The pipeline can use public data available from the \gls{gwosc}~\cite{gwosc}, private data (data stored on the \gls{lvk} servers restricted to members of the collaboration), or local data. %
Data are read using existing {\tt gwpy}~\cite{gwpy} {\tt TimeSeries} methods. %
we denote the raw data measured at detector $I$ over the time period $T$ by $s_I (t_k)$ in what follows, where $t_k$ are discrete times given by $t_k \equiv k \delta t$. %
The values of $k$ are positive integers between 0 and $T/\delta t - 1$ and $\delta t$ is the sampling period, which in \gls{ligo}, Virgo and KAGRA interferometers is 1/(16384 Hz). %
The raw strain data from the two interferometers, $s_I (t_k)$ and $s_J (t_k)$, are downsampled to a user-defined sampling frequency $f_{\rm samp}$, 
using a user-defined re-sampling window (a \textit{Hamming} window by default). %
The downsampling is performed to reduce the memory and computational requirements of the analysis. %
This is achieved using an existing {\tt gwpy TimeSeries} filtering method for strain data. %
Note that selecting an $f_{\rm samp}$ implies fixing a Nyquist frequency of $ f_{\rm Nyquist} = f_{\rm samp}/2$ for the analysis. %
The Nyquist frequency is the highest frequency included in the Fourier expansion at a given sampling rate. Hence, frequencies above it cannot be probed. 
To avoid this becoming a limitation, $f_{\rm samp}$ should be chosen high enough to contain the full spectrum of the signal of interest, within reasonable sensitivity of the detector. %

\begin{figure}
    \centering
    \includegraphics[width=0.55\linewidth]{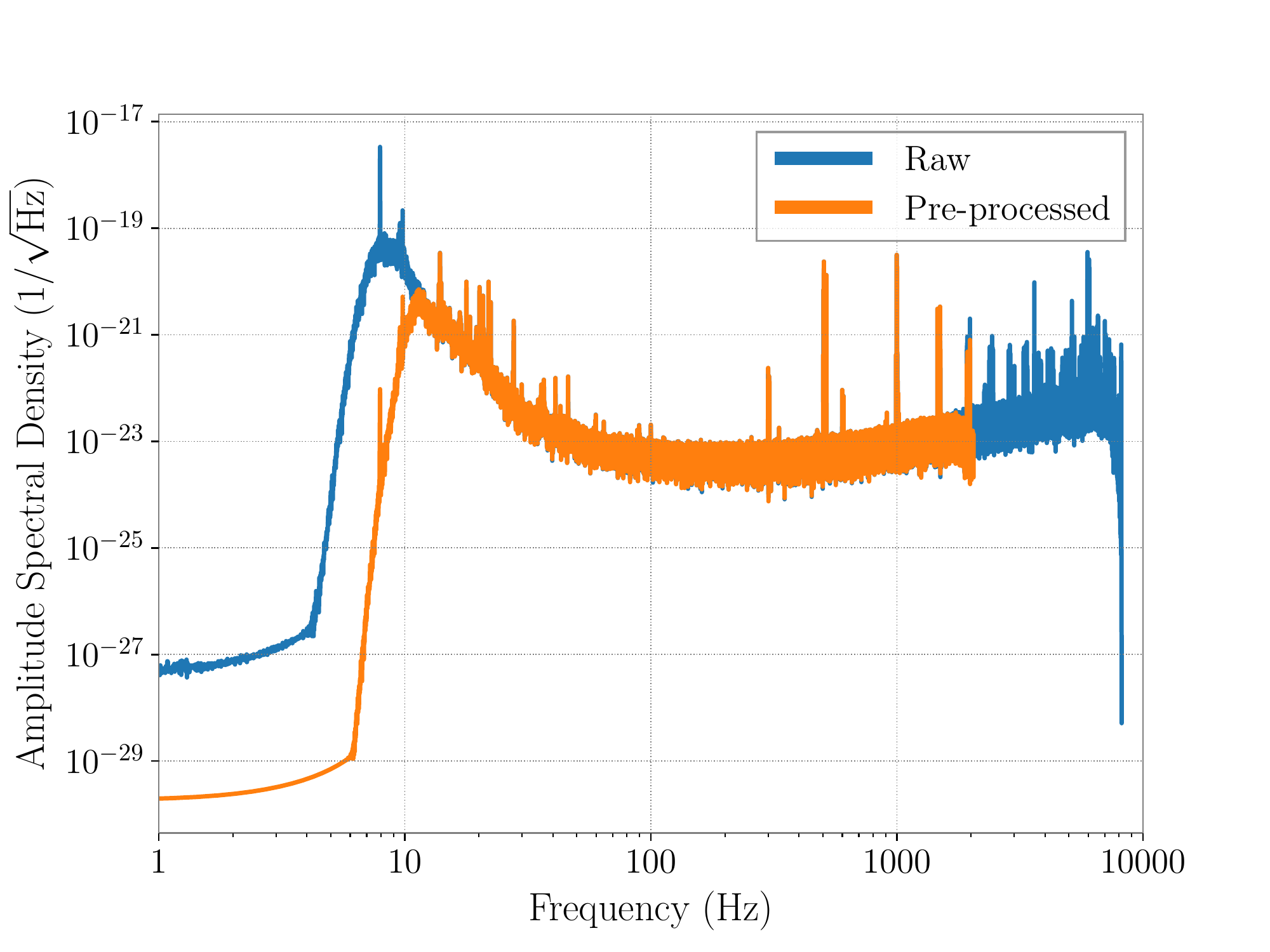}
    \caption{Comparison between the amplitude spectral density (ASD) of a raw (blue solid line) and pre-processed (orange solid line) 192 s segment of \gls{ligo} Livingston O3 data. Pre-processing consists of downsampling the data to 4096 Hz and then removing the low frequency content below 10 Hz.}
    \label{fig:downsampled_highpassed}
\end{figure}

The low-frequency content of ground-based interferometer data (in particular below 10Hz) is 
dominated by seismic and control noise~\cite{PhysRevD.102.062003}. %
For this reason, frequencies below a given (user-defined) cutoff frequency are high-pass filtered, i.e., excluded from the analysis. 
In previous isotropic \gls{gwb} searches~\cite{LIGO_S5, LIGO_O1, LIGO_O2, LIGO_O3}, the input data are high-pass filtered  using a $16$th-order Butterworth filter with a specific knee frequency. %
A 16th-order Butterworth filter is built by first computing its transfer function (in zero-pole-gain form) using the {\tt scipy} library and then filtering the data with the relevant {\tt gwpy TimeSeries} method. The design of the high-pass filter is fixed in the module, only allowing the user to specify the knee frequency. %
The default value of the knee frequency is $11$ Hz, which  was chosen to avoid the spectral leakage from the noise power spectrum below $20$ Hz~\cite{LIGO_O3}. 
See Fig.~\ref{fig:downsampled_highpassed} for an example of data before and after pre-processing. %

At this point, the data may also be screened for large bursts of power in the detector data with high SNR, or \textit{glitches}, due to instrumental or environmental disturbances, which are known to bias estimates of stochastic analyses~\cite{Usman:2015kfa,Pankow:2018qpo,LIGO:2021ppb,Virgo:2022kwz,Davis:2022dnd}. %
Historically, segments with loud glitches were flagged and excluded from analysis by non-stationarity cuts (see Sec.~\ref{Sec:DeltaSigma}). %
In O3, a series of exceptionally loud glitches appeared in the data that led to large fractions of data being removed by previously employed non-stationarity cuts~\cite{LIGO_O3}. %
Hence, an alternative technique called \textit{gating} was employed to address these loud glitches, and drastically reduce the amount of data removed \cite{gatingStudy}. %
Gating is performed internally by {\tt pygwb} by multiplying the data by an inverse {\it Planck-taper} window~\cite{McKechan:2010kp}.
Time periods around samples in the whitened data that have an absolute value above a chosen threshold are marked for gating independently for each interferometer. %
The width of the gate must be sufficiently large to remove the entirety of the relevant glitch. The required width may change based on the data quality of the specific data in the analysis and hence must be empirically determined. %
The tapering length of the window must also be sufficiently long to minimize the addition of artifacts by the gating; 0.25 seconds is found to be sufficient~\cite{Davis:2022dnd}. %
This technique is generically beneficial for the analysis of data that are non-Gaussian, such as real gravitational-wave detector data. %
Gating implemented in {\tt pygwb} is highly customizable to the specific needs of the analysis; default gating parameters are shown below in Table~\ref{tab:parameters}. %
For more details on gating and parameter choices see~\cite{Davis:2022dnd}. %

Finally, the module also allows to perform a time-shifted analysis in which one of the two timeseries is shifted in time by an integer number of seconds before the cross-correlation is performed. %
This technique is employed as a detector noise characterization tool, since it removes the potential correlation due to a broadband \gls{gwb}, while preserving instrumental correlations with coherence times greater than the applied time shift, like nearly-sinusoidal spectral artifacts from, e.g. electronics~\cite{Covas_2018} \footnote{It is worth noting that this time shift will probably not help identify correlated broadband stochastic noise, such as correlated magnetic noise from Schumann resonances, as this is largely caused by lightning strikes and the correlation between detectors is due to seeing the same stochastic signal in both detectors. This is in contrast to chance coherence between coincident periodic artifacts ({\it lines}) at multiple sites that one can find by implementing time shifts.}. %
The time shift is a user defined parameter which should always be greater than the light travel time between detectors (i.e., 10 ms for the LIGO Hanford and Livingston detectors) and smaller than the segment duration. %
Typically, a time shift of 1s is used. \\


\subsection{\tt spectral}\label{sec:spectral}
The role of the {\tt spectral} module is to compute, for each time segment of duration $T$, the discrete frequency domain quantities $C_{IJ, f}$, $P_{I, f}$ and $P_{J, f}$ used in Eqs.~\eqref{eq:Omega} and \eqref{eq:Variance}. The one-sided cross- and auto-power spectral densities $C_{IJ}$ and $P_I$, respectively, of a single segment are defined as
\begin{equation} \label{eq:csd_psd}
    C_{IJ, f} = \frac{2}{T} \tilde s_{I, f}^* \tilde s_{J, f}\,, \ \ \ \ P_{I,f} = \frac{2}{T} |\tilde s_{I, f}|^2 \ ,
\end{equation}
where $\tilde s_f$ are \glspl{dft} of $s(t_k)$ 
 defined by
\begin{equation}
   \tilde s_{f} \equiv \sum_{t_k=0}^{T-\delta t} s(t_k) \, e^{-i 2 \pi m t_k / T} \,,
\end{equation}
where 
$f=m \delta f$, with $m$ a natural number between 0 and $1/(2\,\delta t \,\delta f)$, and $\delta f$ the desired frequency resolution, chosen such that $1/(2\delta t \delta f)$ is an integer. %

The segmented data are windowed before calculating Fourier transforms to avoid spectral leakage due to discontinuities at the ends of the segments. %
The user may define their own choice of window, which defaults to the \textit{Hann} window if none is selected. %
The {\tt spectral} module uses methods from {\tt scipy.signal} to calculate spectrograms $\tilde s^t_{f}$ of the given data, which are then used to calculate the list of $C^t_{IJ, f}$, $P^t_{I, f}$, and $P^t_{J, f}$ quantities, corresponding to different time segments labelled by $t$ in the dataset. %
By default, these are calculated with a 50\% time overlap to account for the impact of the windowing. However, the user may redefine the overlap between consecutive segments to be used throughout the analysis to better suit any choice of window. 

Different averaging procedures are employed to reduce the fluctuations in the spectra estimates and compress the data. %
The procedures we employ are selected to minimize sensitivity loss. %
In the estimates of $P^t_{I, f}$ and $P^t_{J, f}$ 
we employ Welch's estimation method of \glspl{psd}~\cite{pwelch}, which is known to produce minimum variance estimates of the \gls{psd}, implemented as follows. %
Each segment is divided into sub-segments of duration $1/\delta f$ which are \gls{dft}ed individually. %
The auto-correlated power $|\tilde s_{I,f}|^2$ is then averaged over the sub-segments 
to obtain estimates of $P^t_{I, f}$ and $P^t_{J, f}$ for time $t$. %
This procedure returns spectra at the desired frequency resolution $\delta f$, which is typically much larger than the original resolution $1/T$ Hz. %

As the power varies slowly with frequency\footnote{The power varies slowly with frequency except in very few bins, where narrow-band spectral artifacts or {\it lines} are present, as discussed in Sec.~\ref{sec:notch}.}, we can average over neighboring frequencies using a process known as coarse-graining~\cite{https://doi.org/10.48550/arxiv.2106.13785}. %
This is the default procedure employed in the \gls{csd} estimation. %
The resulting spectra are returned at the desired frequency resolution $\delta f$. 
Note that the data are zero-padded before calculating Fourier transforms for $C^t_{IJ, f}$ to avoid wrap-around problems arising from finite data \cite{LIGO_S1, whelan_CC_dcc, numerical_recipes2007}, and hence coarse-graining is required to achieve the desired frequency resolution. %
Zero-padding simply entails appending a vector of zeros equal to the length of the segment before taking the Fourier transform. %

\begin{figure}
    \centering
    \includegraphics[width=0.5\textwidth]{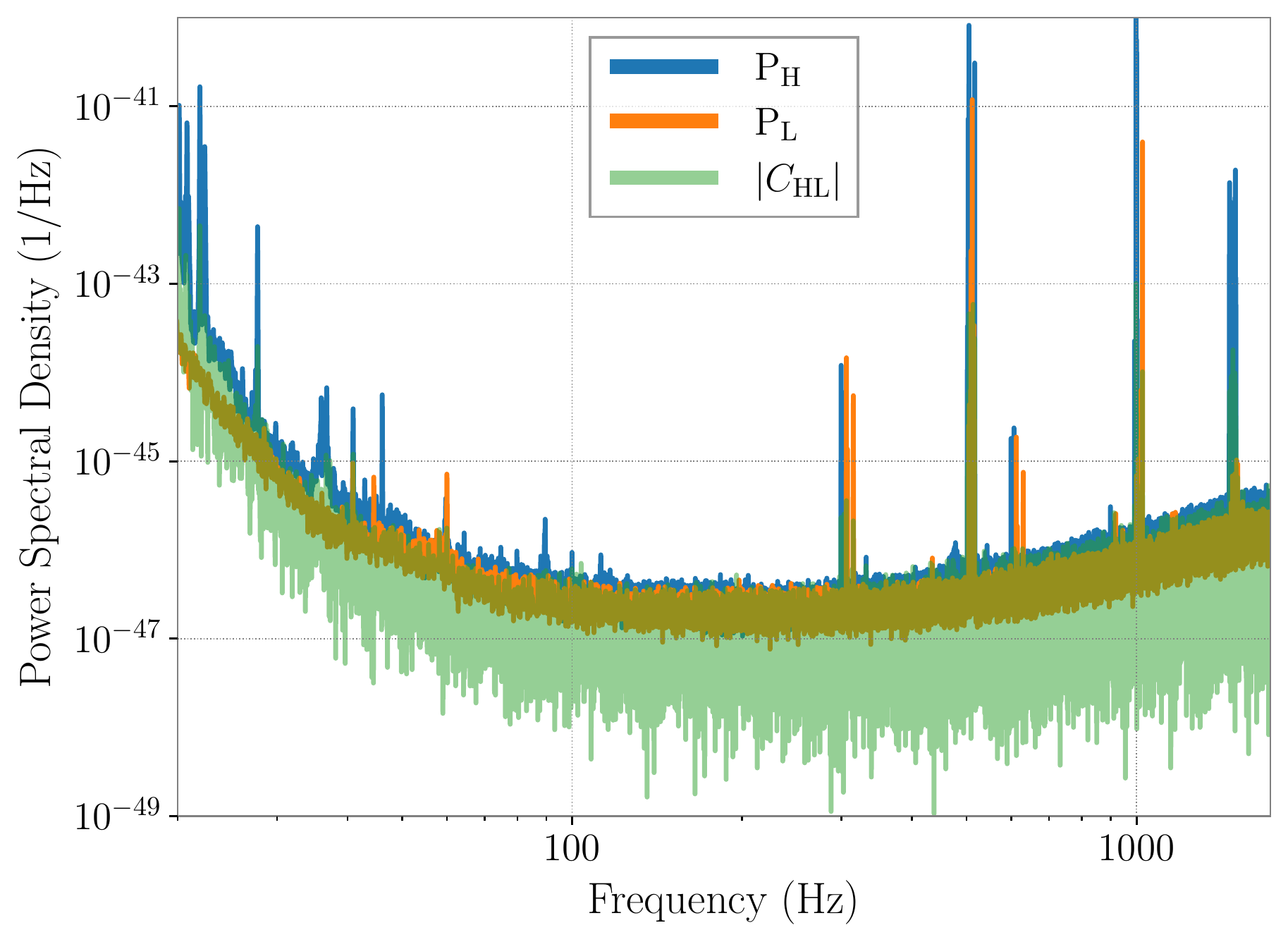}
    \caption{An example of cross- and auto-power spectral densities of the Hanford and Livingston detector data during O3.}
    \label{fig:psd_csd_spectra}
\end{figure}

\begin{figure}
    \centering
    \includegraphics[width=0.7\textwidth]{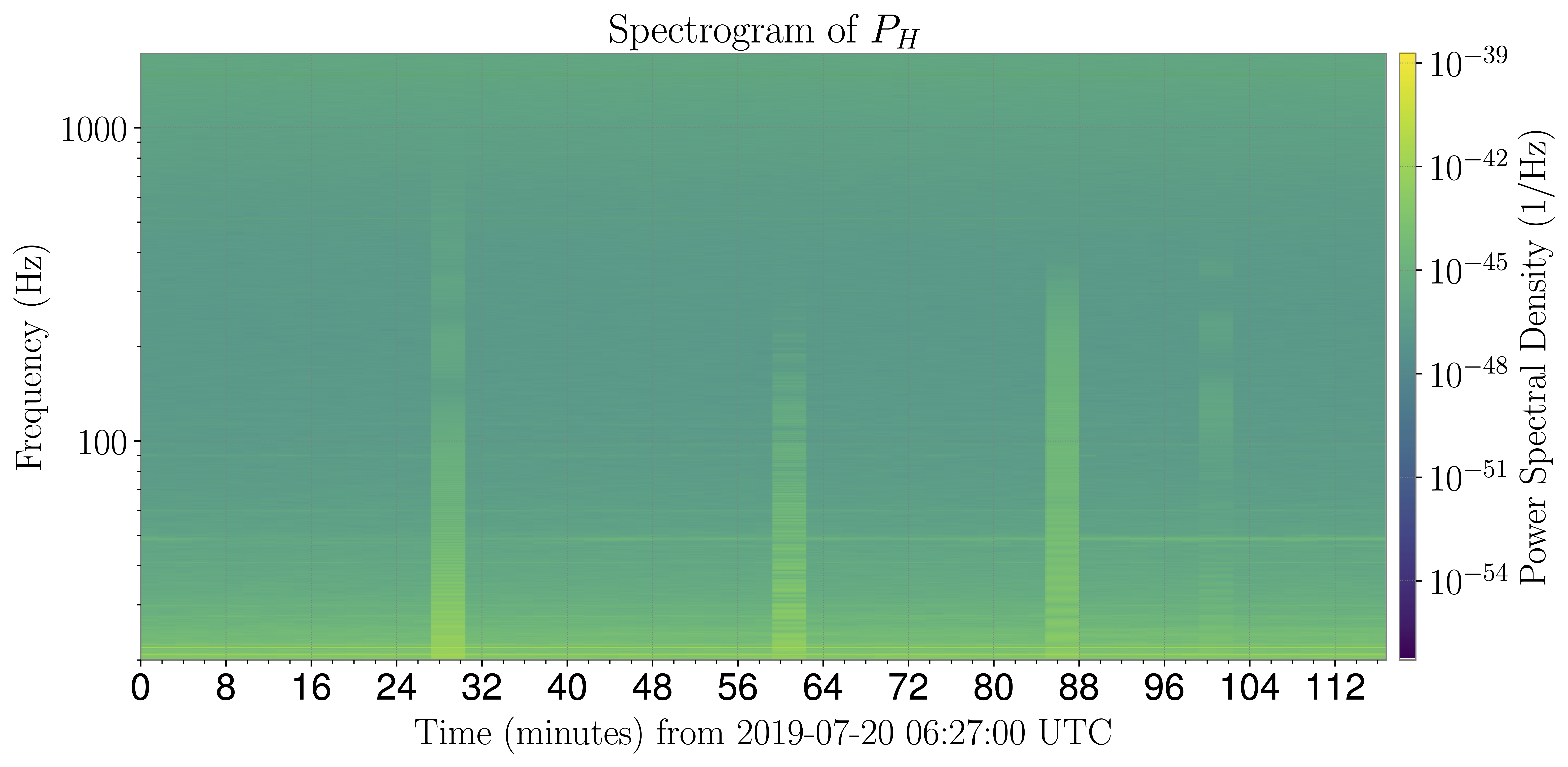}
    \caption{An example spectrogram showing two hours of LIGO   Hanford data during O3. The visible vertical columns correspond to noisy segments, which are usually removed from the analysis (see Sec.~\ref{Sec:DeltaSigma}).}
    \label{fig:psd_spectrogram}
\end{figure}
To further reduce fluctuations in the \gls{psd} estimates, the $P^t_{I, f}$ quantities are averaged over neighboring segments to obtain the final estimate $\bar{P}^t_{I, f}$ of the \gls{psd} at a given time $t$. %
This is appropriate as the noise in \gls{gw} detectors is (most often) approximately stationary over periods of a few minutes. %
We often refer to the initial (un-barred) quantities as ``naive'' and the final (barred) quantities as ``average'' estimates in the rest of this paper, to avoid confusion. %
By default, only nearest neighbors are used for the calculation, such that the \gls{psd} at time $t$ is an average of the naive \glspl{psd} calculated for times $t-T$ and $t+T$. %
The user may define any even number $D$ of segments to be used to perform this average, which are taken before and after the reference time $t$ such that the \gls{psd} is averaged over naive \glspl{psd} at times $t-D T/2$ and $t+DT/2$. 

Fig.~\ref{fig:psd_csd_spectra} shows the cross- and auto-power \glspl{psd} of 192 s of data from the Hanford and Livingston detectors during O3, while Fig.~\ref{fig:psd_spectrogram} shows a two-hour spectrogram of Hanford data during O3, produced with the {\tt spectral} module. %

\subsection{\tt postprocessing}\label{sec:postproc}
Once a set of data, comprised of an uninterrupted stretch of timeseries data, has been pre-processed and average \gls{psd} and \gls{csd} estimates have been calculated for each segment of data within the set, one can combine those separate time segments to construct a final, time-averaged estimate of the \gls{gwb} amplitude. %

Due to the aggressive windowing choice we typically make, and the subsequent overlapping of time segments, we must be careful in combining time segments together. %
The overlapping and windowing cause overlapping time segments to be correlated with one another. %
Within each processed set, individual time segments must be combined while accounting for this covariance. %
A detailed calculation and discussion of this covariance can be found in~\cite{lazz_romano_windowing_note}, while effective approximations to that full calculation can also be used (see, e.g. Sec. IIIB of~\cite{AinDalvi2015}). %

To start, we construct the estimate of the GWB in a single segment $t$. %
As detailed in Sec.~\ref{sec: GWB analysis}, the \gls{gwb} search is often framed in terms of constructing a point estimate for $\Omega_{\rm GW}(f_{\rm ref})$, the energy density of the \gls{gwb} at the specific frequency $f_{\textrm{ref}}$, assuming a power-law for the \gls{gwb} with spectral index $\alpha$. %
We refer to the estimator of this quantity as $\hat \Omega^\alpha_{\mathrm{ref}},$ in general, and for a single time segment of data, it can be constructed using a weighted average over the individual frequency bin estimators $\hat{\Omega}_f$ and ${\sigma}_f$ described in Eqs.~\eqref{eq:Omega} and~\eqref{eq:Variance} calculated per segment $t$, as
\begin{align}
\hat\Omega^\alpha_{\textrm{ref}, t} = \frac{\sum_{f} \hat\Omega_{t,f} H_{{\rm ref}, \alpha}(f)\bar\sigma^{-2}_{t, f}}{\sum_f H_{{\rm ref}, \alpha}^2(f) \bar\sigma^{-2}_{t,f}},\label{eq:omega_alpha}\\
\sigma^\alpha_{\textrm{ref},t} = \left[\sum_f H^2_{{\rm ref}, \alpha}(f)\bar\sigma_{t,f}^{-2}\right]^{-\frac{1}{2}},\label{eq:sigma_alpha}
\end{align}
where the rescaling $H_{{\rm ref}, \alpha}(f)$ is defined in Eq.~\eqref{eq:Salpha}. %
The average variance spectrum per segment, $\bar\sigma^2_{t, f}$, is calculated using average \glspl{psd} described in Sec.~\ref{sec:spectral}. %
These broadband quantities can be calculated for each time segment $t$, and then this set of estimators at each time can be combined to account for the overlap between time segments discussed above. %
We first lay out how to perform this combination assuming we have calculated the quantities above for each individual time segment. %
Then, we discuss how to alternatively average the estimators in each frequency bin over time independently, before combining them into an integrated quantity at the end. %
The latter calculation is normalized such that it gives the same result as the former. %
To avoid heavy notation we drop the bars that indicate average quantities in the rest of this section -- all variances used for the following calculations are average variances as defined above. %

To construct an estimator for the \gls{gwb} using a set of measurements in short, overlapping time segments, we first combine the segments that are non-overlapping. %
If the overlap between segments is $50\%$ or less, then this amounts to separately performing inverse-noise-weighted averaging over the even- and odd-indexed segments:
\begin{align}
\label{eq:odd_even_sigma}
    \sigma_{\mathrm{odd}}^2 &= \frac{1}{\sum_{t\in \mathrm{odd}}\sigma_t^{-2}}\\
    \label{eq:odd_even_omega}
    \Omega_{\textrm{odd}} &= \frac{\sum_{t\in \textrm{odd}}\Omega_t\sigma_{t}^{-2}}{\sum_{t\in\textrm{odd}}\sigma_{t}^{-2}},
\end{align}
where 
the quantities $\Omega_t\equiv\hat\Omega^\alpha_{\textrm{ref}, t}$ and $\sigma_t\equiv\sigma^\alpha_{\textrm{ref}, t}$ for each time segment $t$. %
Analogous expressions are calculated for $\Omega_{\textrm{even}}$ and $\sigma_{\textrm{even}}$. %
Subscripts refer to even/odd time segments, and we drop here the subscripts $_{\rm GW}$, $_{\rm ref}$, and $_\alpha$ used to construct the integrated quantities to lighten the notation. %
We refer to the final, frequency- and time-averaged estimate as $\hat\Omega_\mathrm{ref}$ for now.

Next, we calculate the cross-covariance between point estimates in the odd and even segment combinations~\cite{lazz_romano_windowing_note},
\begin{align}
\sigma_{oe}^2 =\sigma_{eo}^2 &\equiv \langle \Omega_\mathrm{odd}\Omega_\mathrm{even}\rangle - \langle \Omega_\mathrm{odd}\rangle\langle \Omega_\mathrm{even}\rangle\\
&= \frac{1}{2}\frac{\bar{w}^4_\mathrm{ovl}}{\bar{w}^4}\left[\sigma_{\mathrm{odd}}^2 + \sigma_{\mathrm{even}}^2-\frac{1}{2}\sigma_{\mathrm{odd}}^2\sigma_{\mathrm{even}}^2\left(\sigma_{1}^{-2} + \sigma^{-2}_{2M-1}\right)\right],
\end{align}
where $M$ is the number of \textit{independent} segments and so $2M-1$ is the total number of overlapping segments, with the {\it window factors} $\bar{w}^4_\mathrm{ovl}$ and $\bar{w}^4$ as defined in App.~\ref{sec:app_window}. %
For the sake of compactness, we rewrite this as
\begin{align}
    \sigma_{oe}^2 &= \frac{k}{2}\sigma_{\mathrm{odd}}^2\sigma_{\mathrm{even}}^2 \sigma_{IJ}^{-2}\,,\\
    \sigma_{IJ}^2 &= \left[\sigma_{\mathrm{odd}}^{-2} 
 + \sigma_{\mathrm{even}}^{-2} - \frac{1}{2}\left(\sigma_{1}^{-2} + \sigma^{-2}_{2M-1}\right)\right]^{-1}\,,
\end{align}
where $k = \bar{w}^4_\mathrm{ovl} / \bar{w}^4$.

The covariance matrix between even/odd segment sets is then defined as
\begin{align}
    \bm{C} = \begin{pmatrix}\sigma_{\mathrm{odd}}^2 & \sigma_{\mathrm{oe}}^2 \\
            \sigma_{\mathrm{oe}}^2 & \sigma_{\mathrm{even}}^2
    \end{pmatrix},
\end{align}
which we use to construct the optimal combination of segments to obtain the point estimate $\hat{\Omega}_{\mathrm{ref}}$ and its variance $\sigma_{\mathrm{ref}}^2$. These are given by:
\begin{align}
    \hat\Omega_{\mathrm{ref}} &= \frac{\sum_{i=1}^2 \lambda_i \Omega_i}{\sum_{j=1}^2 \lambda_j},\label{eq:ptest_postpoc}\\
    \sigma_{\mathrm{ref}}^2 &= b^2_{\rm avg}\qty({\sum_{k=1}^2 \lambda_k})^{-2}\sum_{i=1}^2\sum_{j=1}^2 \lambda_i C_{ij} \lambda_j, \label{eq:sigma_postpoc}
\end{align}
with
\begin{align}
    \lambda_i = \sum_{j=1}^2 \left(\bm C^{-1}\right)_{ij}\,,
    \label{eq:lambda_i_broadband}
\end{align}
where $i,\,j$ indices label odd/even quantities. %
The bias factor $b_{\rm avg}$ which arises due to harsh windowing of the data has been included in Eq.~\eqref{eq:sigma_postpoc}. %
The derivation of the bias factor is described in App.~\ref{sec:app_window}. %
If combining over non-overlapping segments, then $\sigma_{\mathrm{oe}}^2 = 0,$ and this method reduces to the typical inverse-noise-weighted average that one would expect. %

The above expressions are for a broadband estimator, but in practice the {\tt postprocessing} module combines over time segments before combining over frequency bins. %
We refer to the estimated narrowband quantities as $\hat\Omega_{\mathrm{ref}, f}$ and $\sigma_{\mathrm{ref}, f}$. This notation indicates that, once a power-law spectral model is applied, the estimate in a frequency bin represents an estimate of the GWB at the reference frequency of the power law, assuming the chosen spectral shape.

We normalize $\hat\Omega_{\mathrm{ref}, f}$ and $\sigma_{\textrm{ref}, f}$ such that, when performing a weighted average over frequency bins \textit{after} combining  overlapping time segments we get the same result as Eqs.~(\ref{eq:ptest_postpoc}) and (\ref{eq:sigma_postpoc})  (which assume construction of a broadband estimator \textit{before} combining overlapping time segments). This results in the following expression for $\sigma_{\textrm{ref}, f}^{-2}$,
\begin{align}
\sigma_{\textrm{ref}, f}^{-2} &= b^{-2}_{\rm avg}\frac{\left[\sigma_{\mathrm{odd}, f}^{-2} + \sigma_{\mathrm{even}, f}^{-2} - k\sigma_{IJ, f}^{-2}\right]}{1 - \frac{k^2}{4}\sigma_{\mathrm{odd}}^2\sigma_{\mathrm{even}}^2\sigma_{IJ}^{-4}}\,,
\end{align}
and a corresponding expression for $\hat\Omega_{\mathrm{ref}, f}$,
\begin{align}
\hat\Omega_{\textrm{ref}, f} &= \frac{\Omega^{}_{\textrm{odd}, f}\sigma_{\mathrm{odd}, f}^{-2}\left(1 - \frac{k}{2}\sigma_{\mathrm{odd}}^2\sigma_{IJ}^{-2}\right) + \Omega^{}_{\textrm{even}, f}\sigma_{\mathrm{even}, f}^{-2}\left(1 - \frac{k}{2}\sigma_{\mathrm{even}}^2\sigma_{IJ}^{-2}\right)}{\sigma_{\mathrm{odd}, f}^{-2} + \sigma_{\mathrm{even}, f}^{-2} - k\sigma_{IJ, f}^{-2}}\,.
\end{align}
The even and odd estimators for each frequency bin are defined as in Eqs. (\ref{eq:odd_even_sigma}) and (\ref{eq:odd_even_omega}), except applied to individual bin-by-bin estimators calculated at each time segment. As discussed above, these expressions have been normalized such that 
\begin{align}
   \hat\Omega_{\textrm{ref}} = \frac{\sum_{f}\hat\Omega_{\textrm{ref}, f} \sigma_{\textrm{ref}, f}^{-2}}{\sum_f \sigma_{\textrm{ref}, f}^{-2}}\,, \\ 
   \sigma_{\textrm{ref}}^{2} = \left[\sum_f \sigma_{\textrm{ref}, f}^{-2}\right]^{-1}\,.
\end{align}

The \texttt{postprocessing} module implements the above expressions to estimate $\hat\Omega_{\mathrm{ref}, f}$ and $\sigma_{\mathrm{ref}, f}$ at a fixed $\alpha$, and returns them in form of an \texttt{OmegaSpectrum} object, which sub-classes the classic {\tt gwpy.FrequencySeries} and adds two key attributes: the spectral index $\alpha$ and the reference frequency $f_{\rm ref}$ at which the spectrum is calculated. %
By default, \texttt{pygwb} assumes a power-law spectral index $\alpha=0$ and a reference frequency $f_{\rm ref} = 25$ Hz when constructing the above estimators. %
To explicitly include the $\alpha$ dependence in our results, we refer to the final postprocessed spectra as $\hat\Omega^\alpha_{\mathrm{ref}, f}$ and $\sigma^\alpha_{\mathrm{ref}, f}$.

One of the advantages of averaging over time before averaging over frequency is that one can reweight $\hat\Omega^\alpha_{\mathrm{ref}, f}$ and $\sigma^\alpha_{\mathrm{ref}, f}$ to be estimators for different choices of $\alpha$ without needing to average over all time segments again for a new choice of $\alpha$. %
The \texttt{OmegaSpectrum} object has a built-in method to perform a reweighting to change either $f_\mathrm{ref}$ or $\alpha$ used to calculate $\Omega_\mathrm{ref}$, employing the relation %
\begin{equation}
    \Omega^{{\rm ref}_1, \alpha_1}_{\rm GW}(f) = \Omega^{{\rm ref}_2, \alpha_2}_{\rm GW}(f) \frac{H_{{\rm ref}_1, \alpha_1}(f)}{H_{{\rm ref}_1, \alpha_2}(f)}\,,    
\end{equation}
derived using Eq.~\eqref{eq:omega-plaw}, which implies the following relation between amplitudes at different reference frequencies,
\begin{equation}
    \Omega_{\rm ref_1} = \Omega_{\rm ref_2} \frac{H_{{\rm ref}_2, \alpha}(f)}{H_{{\rm ref}_1, \alpha}(f)}\,.
\end{equation}
This allows to quickly calculate time- and frequency-averaged estimates of the \gls{gwb} amplitude associated with a specific power-law model. %

The default Hubble constant $H_0$, required in the scaling $S_0(f)$ in Eq.~\eqref{eq:S0}, is chosen to be $H_0 = 67.7$~km/(Mpc$\cdot$s), drawn from the Planck 2018 observations~\cite{Planck2018} and imported directly from the {\tt astropy} package. %
This is an attribute of the \texttt{OmegaSpectrum} and may be re-set by the user.

\subsection{\tt delta-sigma cut}\label{Sec:DeltaSigma}
In general, the noise level in ground-based detectors changes slowly on time-scales of tens of minutes to hours. 
The variance $\sigma^2_{\rm GW}$~(see Eq.~\eqref{eq:Variance}) associated to each segment is an indicator of that level of noise, which typically changes at roughly the percent level from one data segment to the next. %
However, there are occasional very loud disturbances to the detectors, such as glitches, which violate the Gaussianity of the noise. %
Auto-gating procedures are in place, as explained in Sec.~\ref{sec:preproc}, to remove loud glitches from the data; however the procedure does not remove all non-stationarities. %
To avoid biases due to these noise events, an automated technique to exclude them from the analysis has been developed~\cite{LIGO_S4}. %
To this end, the {\tt pygwb} package includes the {\tt delta-sigma cut} module, which flags specific segments to be cut from the analyzed set. %
Note that inverse-noise-weighting, as explained in Sec. \ref{sec:postproc}, also reduces the effect of non-Gaussian noise artifacts. %

The ``\textit{$\Delta\sigma$} cut'' calculation consists in comparing the $\sigma_{\rm GW}$ of a segment $t$, $\sigma_t$, to that of its nearest neighbors and flagging it for removal in case their values differ by more than a chosen threshold. %
Conceptually, the calculation is based on the simple inequality,
\begin{equation}
\label{eq:dsc_condition}
    \frac{|\sigma_i - \sigma_{i+1}| + |\sigma_i - \sigma_{i-1}|}{2\sigma_i}>{\rm threshold}\,,
\end{equation}
where $i$ is a segment index. %
However, in practice we perform an analogous, more sophisticated calculation, which compares the naive and average segment variances ${\sigma}_{t, \alpha}$ and $\bar{\sigma}_{t, \alpha}$. %
These are derived from the unweighted naive and average segment variances computed with Eq.~\eqref{eq:Variance} using naive and average \glspl{psd} per segment (see Sec.~\ref{sec:spectral} for details), respectively, which are then reweighted by the index $\alpha$, as shown in Eq.~\eqref{eq:sigma_alpha}. %
The final expression employed in the calculation is 
\begin{equation}
    \frac{|\bar{\sigma}_{t, \alpha} b_{\rm avg} - \sigma_{t, \alpha} b_{\rm nav} |} {\bar{\sigma}_{t, \alpha} b_{\rm avg}}>{\rm threshold}\,,
\end{equation}
which also takes into account the bias factors that arise due to the different impacts of windowing on naive and average quantities (see App.~\ref{sec:app_window} for details). %
Past analyses have used a threshold of 0.2, as this has been shown to yield a Gaussian distribution for the remaining (un-cut) segment variances~\cite{LIGO_S5}. %
For more details on this choice see~\cite{Meyers_thesis}.
\begin{figure}
    \centering
    \includegraphics[width=0.55\linewidth]{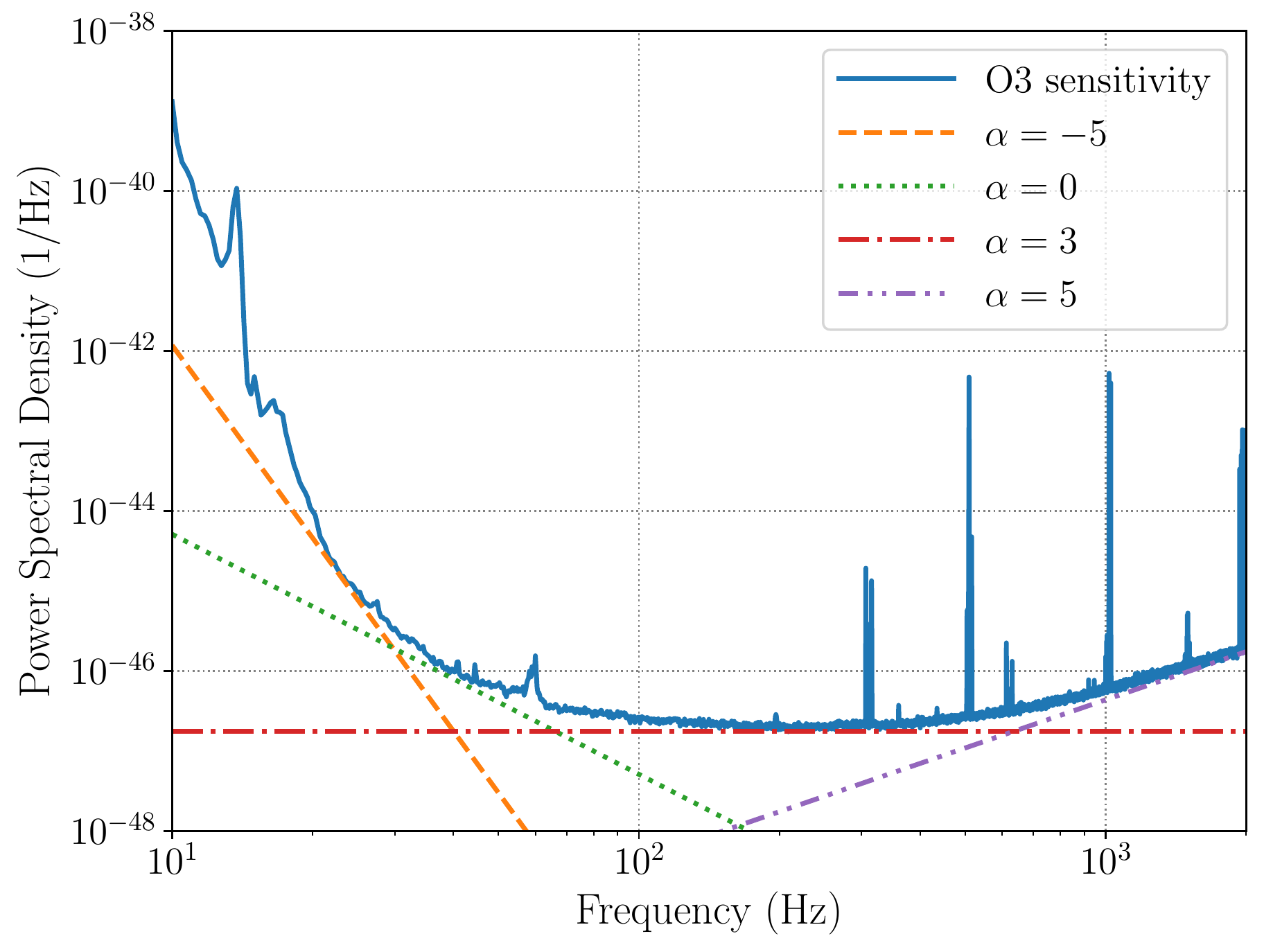}
    \caption{In this plot, power-law spectra with different spectral indices are compared to the O3 sensitivity curve of \gls{ligo}-Livingston. Each power law is sensitive to a different frequency band. This makes it necessary to repeat the \textit{$\Delta\sigma$ cut} assuming different $\alpha$, since this allows to check for noise fluctuations in the whole range of frequencies analyzed. The O3 sensitivity curve for \gls{ligo}-Livingston was retrieved from \cite{O3_sens_curve}.}
    \label{fig:delta_sigma_cut_alphas}
\end{figure}

The $\Delta\sigma$ cut calculation is performed 
assuming different spectral indices $\alpha$ as each power law is sensitive to a different frequency band (see Fig.~\ref{fig:delta_sigma_cut_alphas}). %
The union of all the segments flagged for each $\alpha$ is taken, leading to a full list of segments to discard from the analysis. %
The default choice of $\alpha$ values in the {\tt delta-sigma cut} module is $ \alpha=\{-5, 0, 3\}$, as this adequately covers most of the frequency band of \gls{lvk} searches, from 20-1726 Hz~\cite{LIGO_O3}, at current sensitivity. %
These may be easily modified by the user. %
This would be especially recommended if the search were carried out over a different set of frequencies, or for data from detectors with a spectral sensitivity different than that for Advanced LIGO, Advanced Virgo, or KAGRA. %
Often, the value of $\alpha=5$ is also considered, and was employed in the most recent LVK isotropic search~\cite{O3-iso-KAGRA:2021kbb}. %
The analysis performed at a spectral index $\alpha=-5$ is mostly sensitive to non-stationary effects in the $\sim 15-50$ Hz range, while in the case of $\alpha = 0$ the analysis is sensitive to effects between $\sim 40-80$ Hz, for $\alpha = 3$ from $\sim 90-500$ Hz, and finally $\alpha = 5$ is most sensitive to fluctuations at the higher frequencies, above $\sim 500$ Hz. %
These higher frequencies are not always included in this sort of analysis due to reduced sensitivity in this range, hence $\alpha = 5$ is not a default value used for the cut.


As the \textit{$\Delta\sigma$} cut only compares neighboring segments, long stretches of loud noise--contaminated data can pass the test and be included in the analysis. %
We are currently working to improve this by monitoring and flagging longer stretches of non-stationary noise and prolonged loud noise conditions.

\subsection{\tt notch}\label{sec:notch}

Ground-based laser interferometers present many narrow-frequency noise artifacts which are typically persistent in time, and are generally referred to as noise lines. %
Some examples are calibration lines and mechanical resonances \cite{LIGO:2021ppb,Virgo:2022ysc,vanRemortel_2022}. %
The {\tt notch} module provides the framework to properly deal with these noise lines in the case of the search for an isotropic \gls{gwb}. %
The solution is to ``notch out'' these noise lines, i.e., set the values of the spectra at the affected frequency bins to zero. %
Note that the {\tt notch} module is not built to identify these lines, as this is typically done by detector characterization experts working closely with instrumentalists running the detectors. %
Rather, the final product of the {\tt notch} module is a frequency mask which may be applied to the relevant spectra in the analysis. %

The key object of the {\tt notch} module is the {\tt StochNotchList}, which is a list of {\tt StochNotch} objects. %
A {\tt StochNotch} object represents a physical noise line which has been identified and needs to be removed from the data analysis. 
The object has a minimum and maximum frequency indicating the contaminated frequency region. %
Furthermore, it also comes with a descriptive string which allows the user to keep track of the reason why the line was notched. %
All the different {\tt StochNotch} objects for a certain analysis are then stored in the {\tt StochNotchList} which contains the entire list of lines to be notched from the analysis. %

The notch mask used to apply a set of notches within the analysis is constructed conservatively, such that any frequency that has overlapping frequency content with the noise lines defined in the {\tt StochNotchList} will be removed when applying the notch mask. %
To maintain generality, we discuss here a generic estimated spectrum $\hat\Omega_{f}$, where its value at frequency $f$ estimates $\Omega_{\textrm{GW}}(f)$ in the frequency range [$f - \delta f/2,f + \delta f/2$], where $\delta f$ is the chosen frequency resolution, as defined in Sec.~\ref{sec:spectral}. %
If a noise line has any overlap with the interval [$f - \delta f/2,f + \delta f/2$], the $f$ frequency bin is excluded. %
This implies that a hypothetical delta-peak noise line at $f +\delta f/2$, leads to notching both $f$ as well as $f + \delta f$. %
\begin{figure}
    \centering
    \includegraphics[width = 0.55\textwidth]{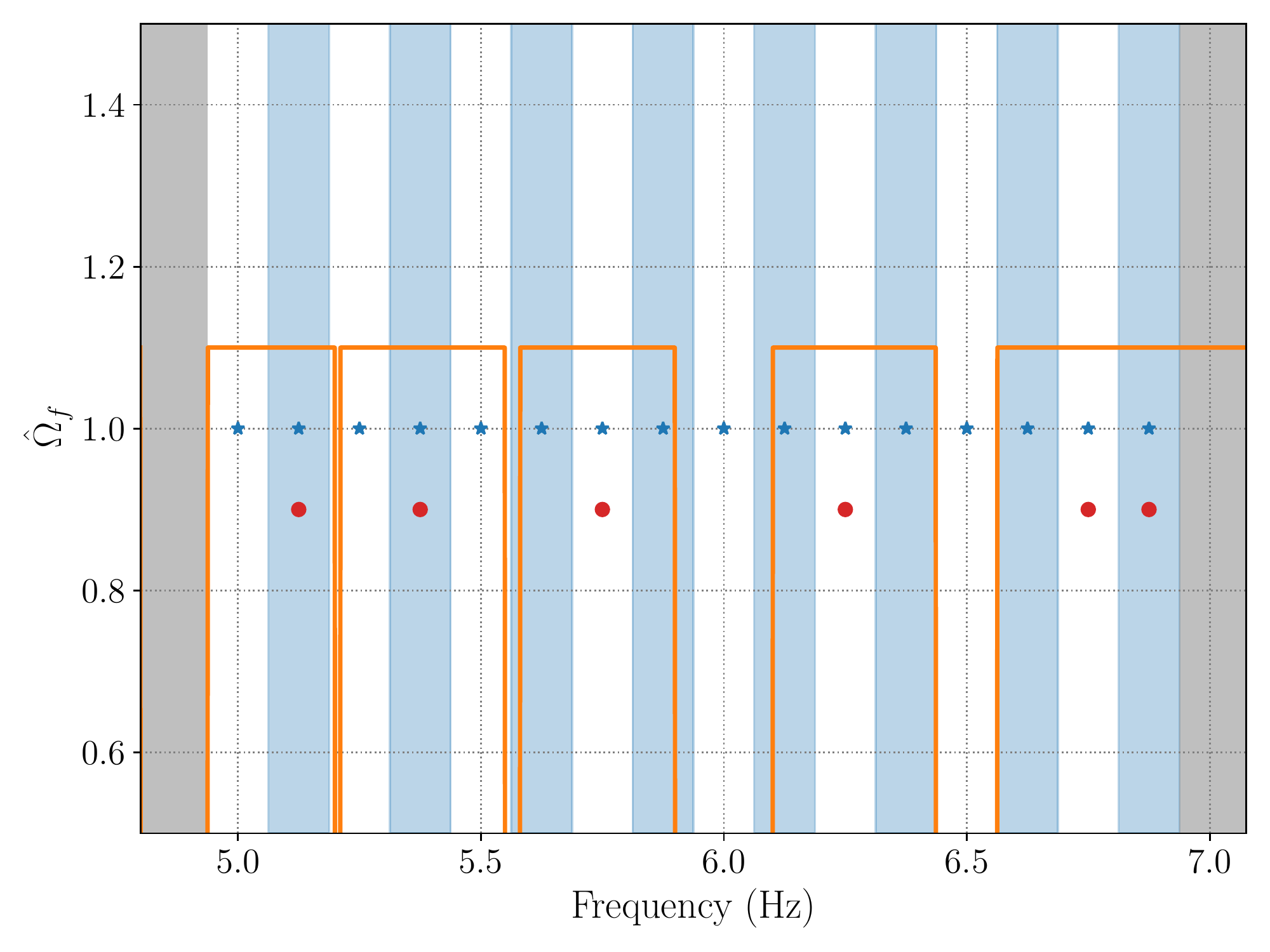}
    \caption{Example of how the notching of noise lines (orange curve) applied to the discrete measurements of the spectrum $\hat\Omega_{\textrm{GW}, f}$ (blue stars) leads to a final set of measurements (red dots). 
    The vertical shaded regions indicate the bins, where even bins are white and odd bins are light blue.    
    The orange line traces out the noise lines such that a noise line is present where the orange curve is zero. 
    The analyzed data spans $[5.0, \,6.875]$ Hz, in the un-shaded region. %
    In this example there are five noise lines, from left to right: a noise line ending at the lowest frequency bin, a noise line entirely contained in one frequency bin, a noise line spread across two frequency bins, a noise line spread across multiple frequency bins, and a noise line from bin-edge to bin-edge. %
    After our notching procedure, the data is reduced to the bins marked by the red dots. %
    For visual convenience we have changed the amplitude in these remaining frequency bins by a factor 0.9. }
    \label{fig:notch_example}
\end{figure}
We present the creation of a notch mask with an example in Fig. \ref{fig:notch_example}, which illustrates how our conservative notching strategy excludes frequency bins based on different scenarios of noise lines. 

The current code is set up to apply the same notches to an entire stretch of data, which can be considered ``time-independent'' notching. %
To allow for time-dependent notching we could either use the current {\tt notch} module and split the analysis in different segments, each having their own notch list. Alternatively, one could extend the current module with an additional parameter which keeps track of which times have to be notched. Since typically the majority of the notched lines in the search for an isotropic \gls{gwb} with data from the \gls{ligo} and Virgo detectors are present during the entire dataset, the possible gain of implementing time-dependent notching is expected to be limited.

\subsection{\tt pe}\label{sec:pe}
Starting from an estimate of the \gls{gwb} spectrum $\hat{\Omega}_{{\rm GW}, f}$, with variance $\sigma^{2}_{{\rm GW}, f}$, it is possible to place stringent constraints on the \gls{gwb} amplitude using a hybrid frequentist-Bayesian approach. %
We consider the general case where we have a set of \gls{gwb} measurements $\hat{\Omega}^{IJ}_{{\rm GW}, f}$ from different detector pairs, or {\it baselines}, $IJ$. %
We define a Gaussian likelihood for $B$ pairs of detectors, 
\begin{equation}
\label{eq:likelihood}
p\qty(\hat{\Omega}^{IJ}_{{\rm GW}, f} | \mathbf{\Theta}) \propto\exp\left[  -\frac{1}{2} \sum_{IJ}^B \sum_f \left(\frac{\hat{\Omega}^{IJ}_{{\rm GW}, f}  - \Omega_{\rm M}(f|\mathbf{\Theta})}{\sigma^{IJ}_{{\rm GW}, f}}\right)^2  \right],
\end{equation}
where $\Omega_{\rm M}(f|\mathbf{\Theta})$ is the \gls{gwb} model and $\mathbf{\Theta}$ are its parameters. %
Bayes' theorem is used to obtain posterior distributions on the model parameters, 
\begin{equation}\label{eq:likelihood_params}
    p\qty(\mathbf{\Theta}|\hat{\Omega}^{IJ}_{{\rm GW}, f}) \propto p\qty(\hat{\Omega}^{IJ}_{{\rm GW}, f}| \mathbf{\Theta})\,p(\mathbf{\Theta})\,,
\end{equation}
where the priors $p(\mathbf{\Theta})$ are employed. %
In practice, when performing parameter estimation on a large dataset, we take the post-processed, {\it unweighted} (i.e., $\alpha=0$) estimate $\hat{\Omega}^{0, IJ}_{{\rm ref}, f}$ to be the measured \gls{gwb} spectrum in each frequency bin, and plug it into Eq.~\eqref{eq:likelihood}. %
Note that it is necessary for the input spectra used in parameter estimation to be unweighted as any other value would constitute a model choice and bias results. %

Within the {\tt pygwb} package, we include the {\tt pe} module to perform parameter estimation as an integral part of the analysis, which naturally follows the computation of the optimal estimate of the \gls{gwb}. %
This is a notable improvement compared to previous LVK analyses, where data products and parameter estimation were handled independently by packages in different programming languages. 
Furthermore, the {\tt pe} module is a simple and user-friendly toolkit for any model builder to constrain their physical models with \gls{gw} data. %

The {\tt pe} module is built on class inheritance, with {\tt GWBModel} as the parent class. %
The methods of the parent class are functions shared between different \gls{gwb} models, e.g., the likelihood formulation in Eq.~(\ref{eq:likelihood}), as well as the noise likelihood, given by Eq.~(\ref{eq:likelihood}) with $\Omega_{\rm M}(f|\mathbf{\Theta})\equiv0$. %
It is possible to include calibration uncertainty by modifying the {\tt calibration\_epsilon} parameter, which defaults to 0. %
For details on the marginalization over calibration uncertainty, see App.~\ref{sec:app_calibration} and \cite{Whelan:2012ur}. %
The \gls{gw} polarization used for analysis is user-defined, and defaults to standard \gls{gr} polarization (i.e., tensor). 
More details on possible polarization choices can be found in Sec.~\ref{sec:baseline}. %
In our implementation of {\tt pe}, we rely on the {\tt Bilby} package~\cite{Ashton:2018jfp} to perform parameter space exploration, and employ the sampler {\tt dynesty} by default \cite{dynesty}. %
The user has flexibility in choosing the sampler as well as the sampler settings. %

Child classes in the {\tt pe} module inherit attributes and methods from the {\tt GWBModel} class. %
Each child class represents a single \gls{gwb} model, and combined they form a catalog of available \gls{gwb} models that may be probed with \gls{gw} data. %
The inheritance structure of the module makes it straightforward to expand the catalog, allowing users of the {\tt pygwb} package to add their own $\Omega_{\rm M}(f|\mathbf{\Theta})$ models. %
The flexibility of the {\tt pe} module allows the user to combine several \gls{gwb} models defined within the module. %
A particularly useful application of this is the modelling of a \gls{gwb} in the presence of correlated magnetic noise, as discussed in \cite{Meyers_2020}, or the simultaneous estimation of astrophysical and cosmological \gls{gwb}s \cite{PhysRevD.103.043023}. %
The {\tt pygwb} documentation~\cite{docs} contains information on the existing models in the catalog, with a description of the GWB models and their parameters. 

\subsection{\tt simulator}
\label{Sec:Simulator}
To both design optimized stochastic analyses and understand our sensitivity to different categories of signals, it is essential to be able to readily simulate realistic interferometer data. %
To this end, the {\tt simulator} module is primarily designed to generate data that corresponds to an isotropic \gls{sgwb} with a given \gls{psd}. %

The \gls{gwb} data in a network of interferometers satisfy a  specific correlation matrix, which includes the set of \glspl{orf} of the entire detector network to account for the spatial separation and relative orientation of the detectors. %
Given a generic signal \gls{psd}, $S_h$, the correlation matrix ${\bm C}(f)$ is given by
\begin{equation}
    C_{IJ}(f) = \delta_{IJ} P_I(f) + \gamma_{IJ}S_h(f). 
\end{equation}
Here $\gamma_{IJ}(f)$ is the normalized \gls{orf} of the baseline $IJ$ as shown in Eq.~\eqref{eq:orf}, hence $\gamma_{II}(f)\equiv 1$, and $P_{I}$ is the noise \gls{psd} of interferometer $I$. %
We have introduced a boldface notation which indicates matrices and vectors which span the detector space. %
The fact that the cross-correlation between detectors for $I\neq J$ only depends on the signal \gls{psd} assumes the noise is uncorrelated across all detectors. %

The simulation of data correlated according to ${\bm C}(f)$ proceeds as follows. %
First, a vector of white, uncorrelated frequency-domain data are generated, ${\bm v}_f$, with a certain frequency resolution $\Delta f$. %
Then, the data are linearly transformed into the correlated ${\bm C}$ space by, 
\begin{equation}
{\bm x}^T_f={\bm v}^T_f~{\sqrt{\bm\Lambda}_f}~{\bm E}^T_f\,,
\end{equation}
where ${\bm \Lambda}$ and ${\bm E}$ are the eigenvalue and eigenvector matrices of $\bm C$, respectively, calculated in each frequency bin. %
This transformation results in data ${\bm x}_f$ that presents the correct correlation, and has been colored with the injected noise and signal power spectra, where appropriate. %
Finally, the frequency-domain data vector is \glspl{ift} to obtain a data vector in the time domain. %

Data generation in the frequency domain, followed by the \gls{ift} to the time domain, can introduce edge-effects in the simulated data segments. 
These may be avoided by \textit{splicing} multiple data segments \cite{TheoryDigitalProcessing}. 
The splicing procedure combines neighboring data segments by windowing and overlapping them, and thus requires more data segments than the actual desired number of segments. 

Concretely, we consider the example where $N_{\rm seg}=1$ and detail the splicing procedure below. %
As the number of desired segments is 1, $2N_{\rm seg}+1=3$ data segments are simulated. %
Assuming these are simulated following the procedure outlined above, we denote these three time-domain data segments by ${\bm x}_0$, ${\bm x}_1$ and ${\bm x}_2$. %
A sine window, defined as
\begin{equation}
    w_j=\sin\left(\frac{\pi j}{N}\right),
\end{equation}
for $0\le j < N$, where $N$ is the number of samples per segment, is used to window the data, 
which are then combined as
\begin{align}
    y_{0j}&= w_j x_{0j},~~~~~~~~~~~~~{\bm z}_0=({\bm y}_0[N/2:N],{\bm{0}})\,,\\
    y_{1j}&= w_j x_{1j},~~~~~~~~~~~~~{\bm z}_1={\bm y}_1\,,\\
    y_{2j}&= w_j x_{2j},~~~~~~~~~~~~~{\bm z}_2=(\mathbf{0},{\bm y}_2[0:N/2])\,,
\end{align}
and finally we obtain a single segment of time domain data $\bf z$,
\begin{equation}
    {\bm z} = {\bm z}_0+{\bm z}_1+{\bm z}_2\,.
\end{equation}
In the above expressions, $\mathbf{0}$ represents an array of zeros with length $N/2$, used to pad the segments, whereas the bracket notation stands for {\tt python} array slicing. %

\begin{figure}
    \centering
\includegraphics[width=.55\textwidth]{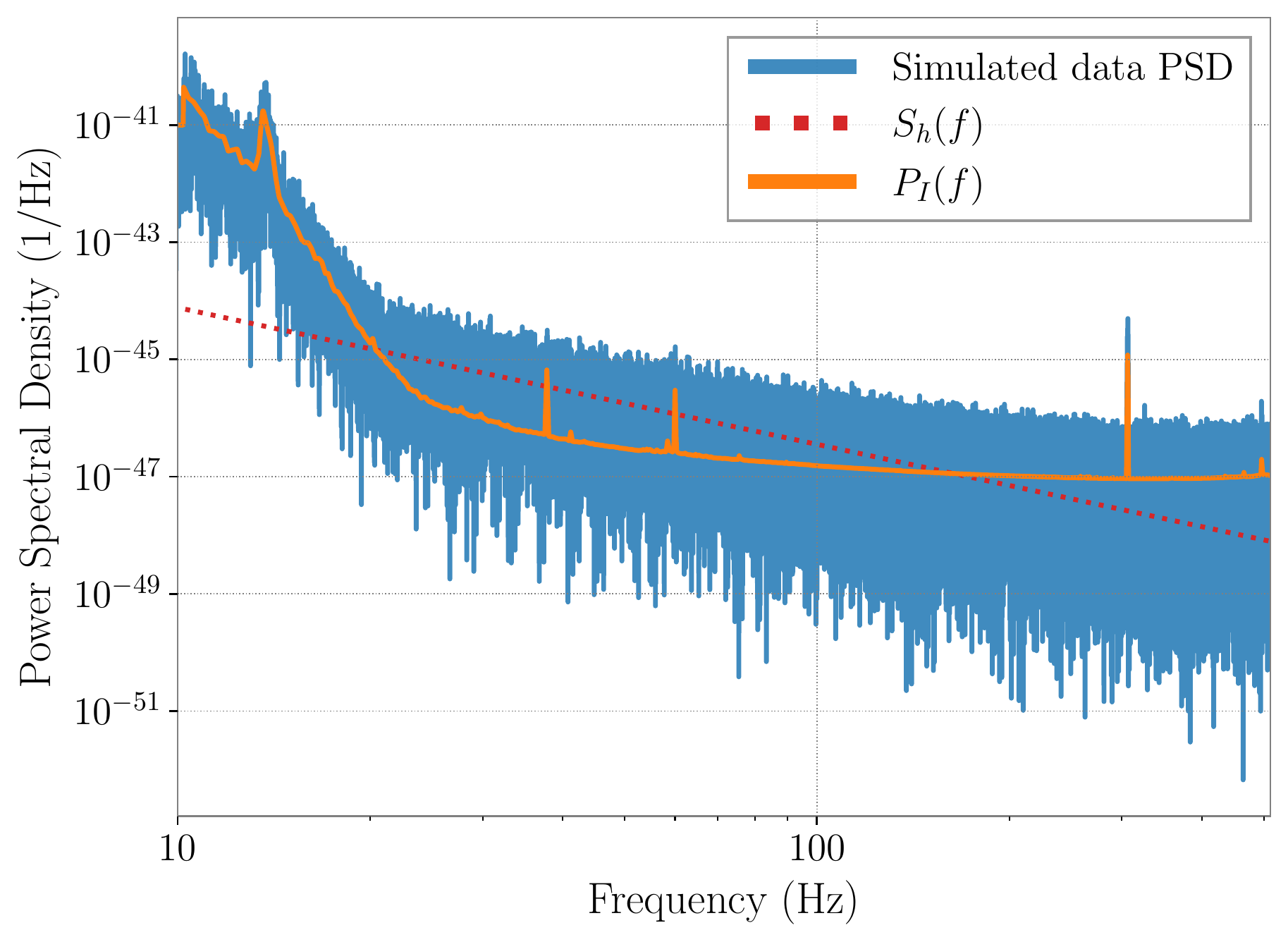}
    \caption{Example of injection using the {\tt simulator} module. The noise \gls{psd} (orange) and the injected signal (red) are clearly discernible as a part of the \gls{psd} of the simulated data (blue).}
    \label{fig:ExampleInjection}
\end{figure}

The splicing procedure can introduce a bias in the simulated power spectrum due to the spectral properties of the window that is applied. %
However, the {\tt simulator} module was tested for several values of the spectral index of a power-law signal \gls{psd}, ranging between $-3$ and 3, yielding a correct injection for these spectral indices. %
The user should nevertheless exercise caution when using the {\tt simulator} module and be aware of the possible introduction of a bias outside the range of tested spectral indices due to splicing.

We show an example of simulated data in Fig.~\ref{fig:ExampleInjection}. %
The injected signal and noise \gls{psd}s are plotted together with the calculated \gls{psd} of a simulated data segment. %
A thorough testing of the {\tt simulator} module is performed in Sec.~\ref{MDCSimulator}. %

\section{Manager objects}
\label{sec: manager objects}
{\tt pygwb} counts three manager objects the user can interface with: {\tt Interferometer}, {\tt Baseline}, and {\tt Network}, which are defined in the {\tt detector}, {\tt baseline}, and {\tt network} modules, respectively. Each object is in charge of storing and saving relevant data, and handles data analysis internally. %
The manager objects are designed such that the user need never call a method from a module directly, but rather will invoke the manager which queries the relevant module to perform the calculation. %
For details on how to use these objects, see the complete set of tutorials in the {\tt pygwb} online documentation~\cite{docs}. %


\subsection{\tt detector}\label{sec:detector}

The \texttt{detector} module is designed to organize the data products related to a \gls{gw} detector and provides functions to create and process its internal data. %
It is formally defined as a subclass of the \texttt{Bilby} \texttt{Interferometer} class~\cite{Ashton:2018jfp}. %
In what follows, we describe the additional features we have developed, and refer the reader to the \texttt{Bilby} documentation for the built-in properties inherited from the parent class. %

By default, the \texttt{Interferometer} object is initialized by taking geometrical information of a \gls{gw} detector such as \texttt{latitude}, \texttt{longitude} and \texttt{elevation} as arguments. %
\begin{Verbatim}[commandchars=\\\{\},frame=leftline,framesep=1.5ex,framerule=0.8pt,fontsize=\small]
\PY{k+kn}{from} \PY{n+nn}{pygwb} \PY{k+kn}{import} \PY{n}{detector}
\PY{n}{my\PYZus{}detector} \PY{o}{=} \PY{n}{detector}\PY{o}{.}\PY{n}{Interferometer}\PY{p}{(}\PY{o}{*}\PY{n}{args}\PY{p}{,} \PY{o}{*}\PY{o}{*}\PY{n}{kwargs}\PY{p}{)}
\end{Verbatim}

While the above method allows the user to customize the detector's specification, one can initialize the object based on existing \gls{gw} detectors as done in \texttt{bilby}'s \texttt{Interferometer} class by calling the {\tt get\_empty\_interferometer} method. %

Once initialized, this object provides several ways to read in and process timeseries data, all of which internally call the \texttt{preprocessing} module, using information such as a channel name to query the data, or pointing to a {\tt numpy} array or a {\tt gwpy} object directly. %
Additionally, the {\tt Interferometer} object includes processing methods relying on the {\tt spectral} module described in Sec.~\ref{sec:spectral} to calculate naive and averaged spectrograms from the stored timeseries data.

A pair of {\tt Interferometer} objects can be used to initialize a {\tt Baseline} object, as described below. These are then imported as attributes of the \texttt{Baseline} object and store data products specific to each detector.

\subsection{\tt baseline}\label{sec:baseline}
The {\tt Baseline} module is by design the core of the {\tt pygwb} stochastic analysis. %
Its main role is to manage the cross-correlation between {\tt Interferometer} data products, combine these into a single cross-spectrum, which represents the point estimate of the analysis, and calculate the associated error, as introduced in Sec.~\ref{sec: GWB analysis}.

The standard initialization of a {\tt Baseline} object simply requires a pair of {\tt Interferometer} objects. %
\begin{Verbatim}[commandchars=\\\{\},frame=leftline,framesep=1.5ex,framerule=0.8pt,fontsize=\small]
\PY{k+kn}{from} \PY{n+nn}{pygwb} \PY{k+kn}{import} \PY{n}{baseline}
\PY{n}{H2H2\PYZus{}baseline} \PY{o}{=} \PY{n}{baseline}\PY{o}{.}\PY{n}{Baseline}\PY{p}{(}\PY{l+s+s2}{\PYZdq{}}\PY{l+s+s2}{H1\PYZhy{}H2}\PY{l+s+s2}{\PYZdq{}}\PY{p}{,} \PY{n}{H1}\PY{p}{,} \PY{n}{H2}\PY{p}{)}
\end{Verbatim}

Here {\tt H1} and {\tt H2} are {\tt Interferometer} objects. %
It is also possible to load a previously stored {\tt Baseline} object in {\tt pickle} format by calling the relevant class method. %

The data loaded into the {\tt Interferometer} objects are automatically imported into the {\tt Baseline} object upon initialization. %
The {\tt Baseline} object relies on the {\tt spectral} module to calculate cross-correlations between the data streams, following the methodology shown in Sec.~\ref{sec:spectral}. %
Similarly, it relies on the {\tt postprocessing} module to obtain the point estimate $\hat{\Omega}^\alpha_{\rm ref}$ and its variance $\sigma^\alpha_{\rm ref}$, as described in Eqs.~(\ref{eq:ptest_postpoc}--\ref{eq:sigma_postpoc}). %
The user may choose to calculate point estimate and sigma spectra or point values; in the latter case, the spectra are automatically stored to facilitate subsequent analyses. %

Calculating $\hat{\Omega}^\alpha_{\rm ref}$, as well as performing parameter estimation on the GWB spectrum, requires the two-detector \gls{orf}, $\gamma_{IJ}$, shown in Eq.~\eqref{eq:orf}. %
The \gls{orf} is calculated at {\tt Baseline} object initialization, then stored as an attribute. %
By default, we assume \gls{gr}, which presents two independent degrees of freedom for the strain field, typically $A=\{+, \times\}$ in the transverse-traceless gauge. For a precise derivation of this function and detector response definitions, see for example~\cite{Romano_2017}. %

The {\tt Baseline} object is also equipped to probe circularly polarized backgrounds~\cite{Seto:2007tn}, and non-\gls{gr} polarizations in the \gls{gwb}, such as scalar and vector backgrounds~\cite{TeVeS}. %
This requires selecting a different choice of $A$, according to the chosen polarization type, which can be declared when calculating $\hat{\Omega}^\alpha_{\rm ref}$ or the \gls{orf} directly. %
Details on the expressions for non-\gls{gr} $\gamma_{IJ}$ functions may be found in the appendix of~\cite{TeVeS}.

\subsection{\tt network}\label{sec:network}

The {\tt network} module is designed to handle two different tasks. %
Its primary purpose is to combine results from different {\tt Baseline} objects. %
Similarly to the {\tt Baseline} object, the {\tt Network} object imports {\tt Baseline} objects as attributes which may be invoked through the {\tt Network}. %
In addition to this functionality, it can also be used to simulate cross-correlated data across a network of detectors. %
Both signal-only and signal and noise data can be simulated using a {\tt Network} object. %
The {\tt network} module handles all data generation by querying the {\tt simulator} module. %

The {\tt Network} object can be initialized in two ways. %
By default it is initialized through a list of {\tt Interferometer} objects. 
\begin{Verbatim}[commandchars=\\\{\},frame=leftline,framesep=1.5ex,framerule=0.8pt,fontsize=\small]
\PY{k+kn}{from} \PY{n+nn}{pygwb} \PY{k+kn}{import} \PY{n}{network}
\PY{n}{HLV\PYZus{}network} \PY{o}{=} \PY{n}{network}\PY{o}{.}\PY{n}{Network}\PY{p}{(}\PY{l+s+s1}{\PYZsq{}}\PY{l+s+s1}{HLV}\PY{l+s+s1}{\PYZsq{}}\PY{p}{,} \PY{p}{[}\PY{n}{H1}\PY{p}{,} \PY{n}{L1}\PY{p}{,} \PY{n}{V1}\PY{p}{]}\PY{p}{)}
\end{Verbatim}

It is also possible to initialize a {\tt Network} using a list of {\tt Baseline} objects, to streamline the combination of results from different baselines which already contain final data products. 
\begin{Verbatim}[commandchars=\\\{\},frame=leftline,framesep=1.5ex,framerule=0.8pt,fontsize=\small]
\PY{n}{HLV\PYZus{}network} \PY{o}{=} \PY{n}{network}\PY{o}{.}\PY{n}{Network}\PY{o}{.}\PY{n}{from\PYZus{}baselines}\PY{p}{(}\PY{l+s+s1}{\PYZsq{}}\PY{l+s+s1}{HLV}\PY{l+s+s1}{\PYZsq{}}\PY{p}{,} \PY{p}{[}\PY{n}{HL\PYZus{}baseline}\PY{p}{,} \PY{n}{HV\PYZus{}baseline}\PY{p}{,} \PY{n}{LV\PYZus{}baseline}\PY{p}{]}\PY{p}{)}
\end{Verbatim}

The combined point estimate and sigma spectra are stored as attributes of the {\tt Network}. %
These are combined by performing an inverse-noise--weighted average over the individual {Baseline} final spectra, %
assuming the data are uncorrelated between baselines, i.e., assuming each baseline provides independent information. %
This is a valid approximation when working in the large noise limit. %
Further details can be found in \cite{Allen:1999}.

The {\tt Network} is also designed to produce appropriately correlated simulated data for a network of interferometers. %
The {\tt Network} can either simulate data from scratch, or add simulated data to pre-existing data, if the interferometers used to initialize the object contain strain data. %
The latter is simply done as strain adds coherently in the time domain. %
This functionality relies on the {\tt simulator} module which performs the data simulation, as discussed in Sec. \ref{Sec:Simulator}. 

\section{Analysis pipeline}
\label{sec: pipeline}

The previous sections contain a detailed description of each of the modules of the {\tt pygwb} package. %
We now present an overview of the package analysis scripts,  which combine the various modules into a \gls{gwb} analysis pipeline. %
The pipeline has several default values which may be changed according to the user's requirements. %
However, we note that thanks to the flexibility of the {\tt pygwb} package, one can also easily construct an ad-hoc pipeline. %

\subsection{\tt pygwb\_pipe}
\begin{table}[h!]
    \centering
    \footnotesize
    \begin{tabular}{c|c|c}
        Parameter & Default value & Description \\
        \hline
        \hline
        \multicolumn{3}{c}{Script arguments}\\
        \hline
        {\tt output\_path} & {\tt ""} &
                        Output data path \\
        {\tt calc\_pt\_est} & {\tt True} & 
                        If {\tt True}, calculate point estimates \\
        {\tt apply\_dsc} & {\tt True} &
                         If {\tt True}, apply $\Delta\sigma$ cut \\
        {\tt pickle\_out} & {\tt True} &
                        If {\tt True}, pickle post-processed baseline \\
        {\tt wipe\_ifo} & {\tt True} & If {\tt True}, set interferometer strain data to 0\\
        \hline
        \multicolumn{3}{c}{Data specifics}\\
        \hline
        {\tt interferometer\_list} & {\tt ["H1", "L1"]} & List of (2) interferometers \\

        {\tt t0} & 0 & Analysis start time \\
        {\tt tf} & 100 & Analysis end time \\
        {\tt data\_type} & {\tt public} & Data accessibility \\
        {\tt channel} & {\tt GWOSC-16KHZ\_R1\_STRAIN} & Data channel name \\
        \hline
        \multicolumn{3}{c}{Pre-processing}\\
        \hline
        {\tt tag} & C00 & Descriptive data tag \\
        {\tt new\_sample\_rate} & 4096 Hz & Downsampled sample rate \\
        {\tt input\_sample\_rate} & 16384 Hz & Input sample rate \\
        {\tt cutoff\_frequency} & 11 Hz & Lower frequency cutoff \\
        {\tt segment\_duration} & 192 s & Individual segment duration \\
        {\tt number\_cropped\_seconds} & 2 s & Preprocessing cropped seconds \\
        {\tt window\_downsampling} & hamming & Downsampling window \\
        {\tt ftype} & fir & Downsampling filter \\
        {\tt time\_shift} & 0 s & Time shift duration \\
        \hline
        \multicolumn{3}{c}{Gating}\\
        \hline
        {\tt gate\_data} & {\tt False} & If {\tt True}, self-gate data \\
        {\tt gate\_tzero} & 1 s & 0 time half-width duration \\
        {\tt gate\_tpad} & 0.5 s & Gating window tapering \\
        {\tt gate\_threshold} & 50 & Gating threshold \\
        {\tt cluster\_window} & 0.5 & Gating cluster window \\
        {\tt gate\_whiten} & {\tt True} & If {\tt True}, whiten data before gating \\
        \hline
        \multicolumn{3}{c}{Spectral density estimation}\\
        \hline
        {\tt frequency\_resolution} & 1/32 Hz & Output frequency resolution \\
        {\tt overlap\_factor} & 0.5 & Consecutive segment fractional overlap \\
        {\tt N\_average\_segments\_welch\_psd} & 2 & Average \gls{psd} segment number \\
        {\tt zeropad\_csd} & {\tt True} & If true zeropad the \gls{csd}\\
        \hline
        \multicolumn{3}{c}{FFT window specifics}\\
        \hline
        {\tt window\_fft\_dict} & {\tt hann} & FFT window parameter dictionary\\
        \hline
        \multicolumn{3}{c}{Postprocessing}\\
        \hline
        {\tt polarization} & {\tt tensor} & ORF polarization basis \\
        {\tt alpha} & 0 & Spectral index $\alpha$ \\
        {\tt fref} & 25 Hz & Reference frequency $f_{\rm ref}$ \\
        {\tt flow} & 20 Hz & Lowest frequency included \\
        {\tt fhigh} & 1726 Hz & Highest frequency included \\
        \hline
        \multicolumn{3}{c}{Data quality specifics}\\
        \hline
        {\tt notch\_list\_path} & {\tt ""} & Notch list file path \\
        {\tt calibration\_epsilon} & 0 & Calibration coefficient \\
        {\tt alphas\_delta\_sigma\_cut} & [-5, 0, 3] & List of $\Delta\sigma$ cut spectral indices \\
        {\tt delta\_sigma\_cut} & 0.2 & $\Delta\sigma$ cut cutoff value \\
        {\tt return\_naive\_and\_averaged\_sigmas} & {\tt False} & If {\tt True}, return both $\sigma$ and $\bar{\sigma}$  \\
        & & used in $\Delta\sigma$ calculation \\
        \hline
        \multicolumn{3}{c}{Output specifics}\\
        \hline
        {\tt save\_data\_type} & {\tt npz} & Output datatype \\
        \hline
        \multicolumn{3}{c}{Local data locations}\\
        \hline
        {\tt local\_data\_path\_dict} & {\tt\{\}} & Dictionary of local data paths         \end{tabular}
    \caption{Default parameters for the {\tt pygwb$\_$pipe} script as well as the {\tt Parameters} dataclass. Most of these choices reflect defaults chosen in the past when analysing LIGO and Virgo data. Notably, the default start and end times for the analysis are not meaningful and represent placeholders for the user-defined times. A default initialization file is included in the package with meaningful start and end times present in the O3 open dataset.}
    \label{tab:parameters}
\end{table}

The core script of our analysis suite, {\tt pygwb$\_$pipe}, is designed to carry out the bulk of the stochastic analysis. %
It combines the {\tt pygwb} modules in order to go from the unprocessed data to the optimally averaged $\hat{\Omega}^\alpha_{{\rm ref}, f}$ and $\sigma^\alpha_{{\rm ref}, f}$ spectra for a single baseline. %
 To read in the analysis parameters, {\tt pygwb\_pipe} interfaces with the {\tt parameters} module, specifically designed to handle the analysis parameters, either passed through an initialization file ({\tt param\_file}) or declared in the command line. %
 The module includes the {\tt Parameters} dataclass which stores the chosen parameters. %
 The pipeline may be run from the command line as follows.
\begin{Verbatim}[commandchars=\\\{\},frame=leftline,framesep=1.5ex,framerule=0.8pt,fontsize=\small]
\PY{n}{pygwb\PYZus{}pipe} \PY{o}{\PYZhy{}}\PY{o}{\PYZhy{}}\PY{n}{param\PYZus{}file} \PY{p}{\PYZob{}}\PY{n}{path\PYZus{}to\PYZus{}param\PYZus{}file}\PY{p}{\PYZcb{}}
\end{Verbatim}

All {\tt param\_file} parameters may be alternatively passed from the command line directly. %
If a mixture of parameter file and command line parameters are passed, the latter will override their corresponding values stored in the parameter file. %
Additionally, a set of pipeline--specific parameters may be passed from the command line for ease of use, such as whether to apply data quality cuts. %
A full list of parameters and their description may be found in Table \ref{tab:parameters}. %

After reading in the parameters, two {\tt Interferometer} objects are created accordingly, 
and data are loaded in and pre-processed using the {\tt preprocessing} module. %
Depending on the value of the {\tt gate\_data} parameter in the initialization file, the gating outlined in Sec.~\ref{sec:preproc} also takes place at this stage. %
Subsequently, a baseline object is created using the pair of interferometer objects. %
Recall that the {\tt baseline} module plays a central role in the pipeline and handles the computation of the various quantities of interest, including the (average) \gls{psd}s and \gls{csd}s of the baseline, relying on the {\tt spectral} module. %
This is described in more detail in Sec.s~\ref{sec:spectral} and~\ref{sec:baseline}.

The delta-sigma cut is then performed,
and optimally averaged spectra and overall point estimate are calculated with the relevant {\tt Baseline} methods. %
The delta-sigma cut is applied by default, but may also be calculated and applied at a later stage. %
Finally, the spectra, the overall point estimate, and the pickled baselines (if requested), are saved as output. %
By default, the output is in {\tt numpy} binary file format, {\tt npz}. %

In realistic scenarios, we analyze year-long datasets and running {\tt pygwb\_pipe} in series is sub-optimal. %
However, a long dataset can be split into smaller jobs and parallelized on a cluster. %
The output of each job is then combined into a single set of result spectra $\hat\Omega^\alpha_{{\rm ref}, f}$ and $\sigma_{{\rm ref}, f}$ using the {\tt pygwb\_combine} script. %
The latter simply takes a weighted average over all jobs, assuming each job is an independent measurement of the signal. %
At this stage it is possible to implement final post-processing choices, such as re-weighting the spectra to a desired $\alpha$ and $f_{\rm ref}$, as well as change the default Hubble constant $H_0$ at which results are reported. %

Details on running the pipeline and combination scripts may be found in our online documentation~\cite{docs}. %

\subsection{\tt statistical\_checks}\label{Sec:StatChecks}
With the {\tt statistical\_checks} module, we provide a tool to perform initial statistical analyses of a {\tt pygwb} run result set, and visualize them in pre-formatted plots. %
We identify five broad categories of checks. %

The first set calculates the running point estimate for $\hat\Omega_{\rm ref}^\alpha$ and $\sigma_{\rm ref}^\alpha$ quantities as a function of time, as more data segments are added to the analysis. %
The values of $\alpha$ and $f_{\rm ref}$ are those used in the analysis and may not be changed at this point. %
The running averages are cumulative weighted averages of time--ordered segments, and do not take segment-by-segment correlation into account. %
In case of detection, these converge to a biased point estimate and $\sigma$, as proper postprocessing is not applied (see Sec.~\ref{sec:postproc}). %
However, the visualization of running quantities is extremely useful to identify trends in the data, and ultimately will flag a possible detection. %
The module also provides a linear trend analysis, fitting the evolution of the parameters described above as a function of time. %

\begin{figure}[t]
    \centering
    \includegraphics[width=0.47\textwidth]{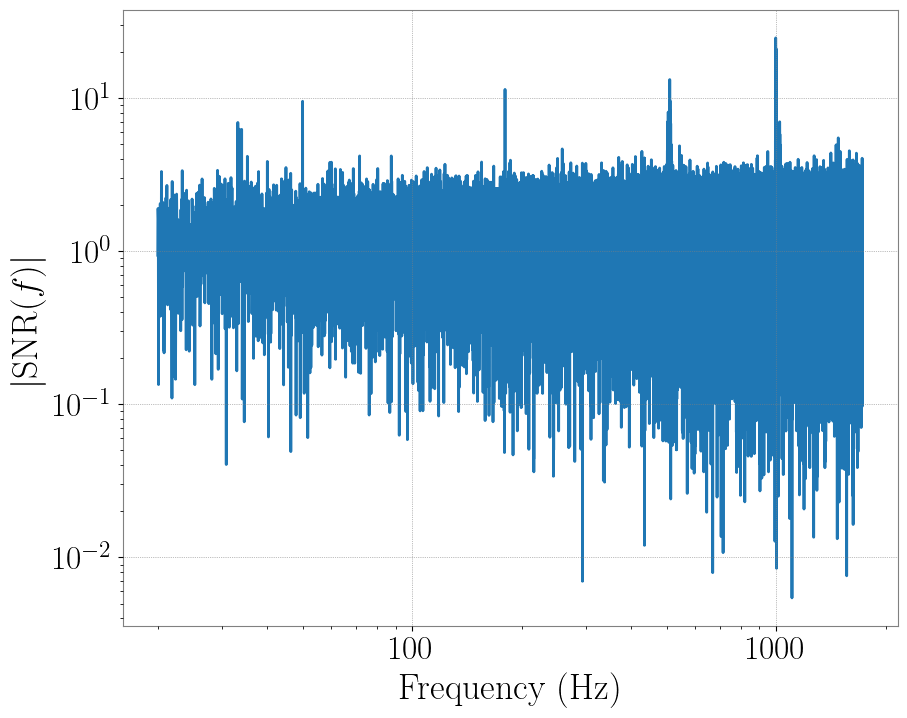}
    \includegraphics[width=0.47\textwidth]{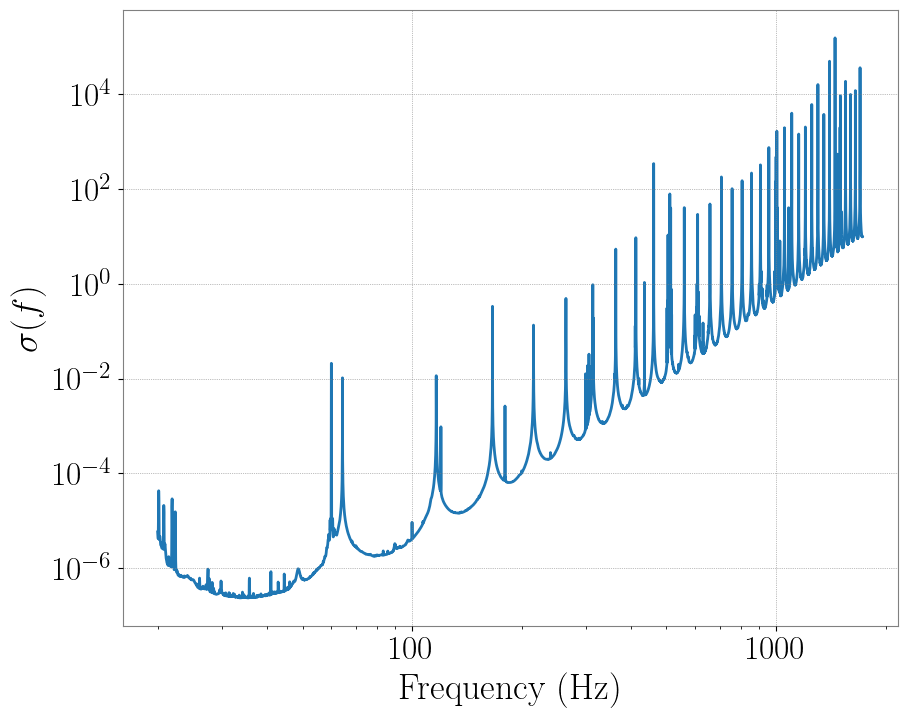}
    \caption{\textbf{Left}: Absolute value of the \gls{snr} spectrum as a function of frequencies. \textbf{Right}: Sigma spectrum as a function of frequency.}
    \label{fig:FreqPlots}
\end{figure}

The second set focuses on the \gls{snr} spectrum as a function of frequency, defined as
\begin{equation}
    {\rm SNR}_f = \frac{\hat\Omega^\alpha_{{\rm ref}, f}}{\sigma^\alpha_{{\rm ref}, f}}\,.
\end{equation}
The absolute value, real, and imaginary part of the \gls{snr} are calculated, as well as the cumulative \gls{snr}. %
An example of these plots using the first sub-set of O3 data further described in Sec. \ref{Sec:O3Data} is given in Fig.~\ref{fig:FreqPlots}. %
These plots are a faithful representation of the ``noisiness'' of each frequency bin and how much each bin contributes to the analysis. %

\begin{figure}[h!]
    \centering
    \includegraphics[width=.43\textwidth]{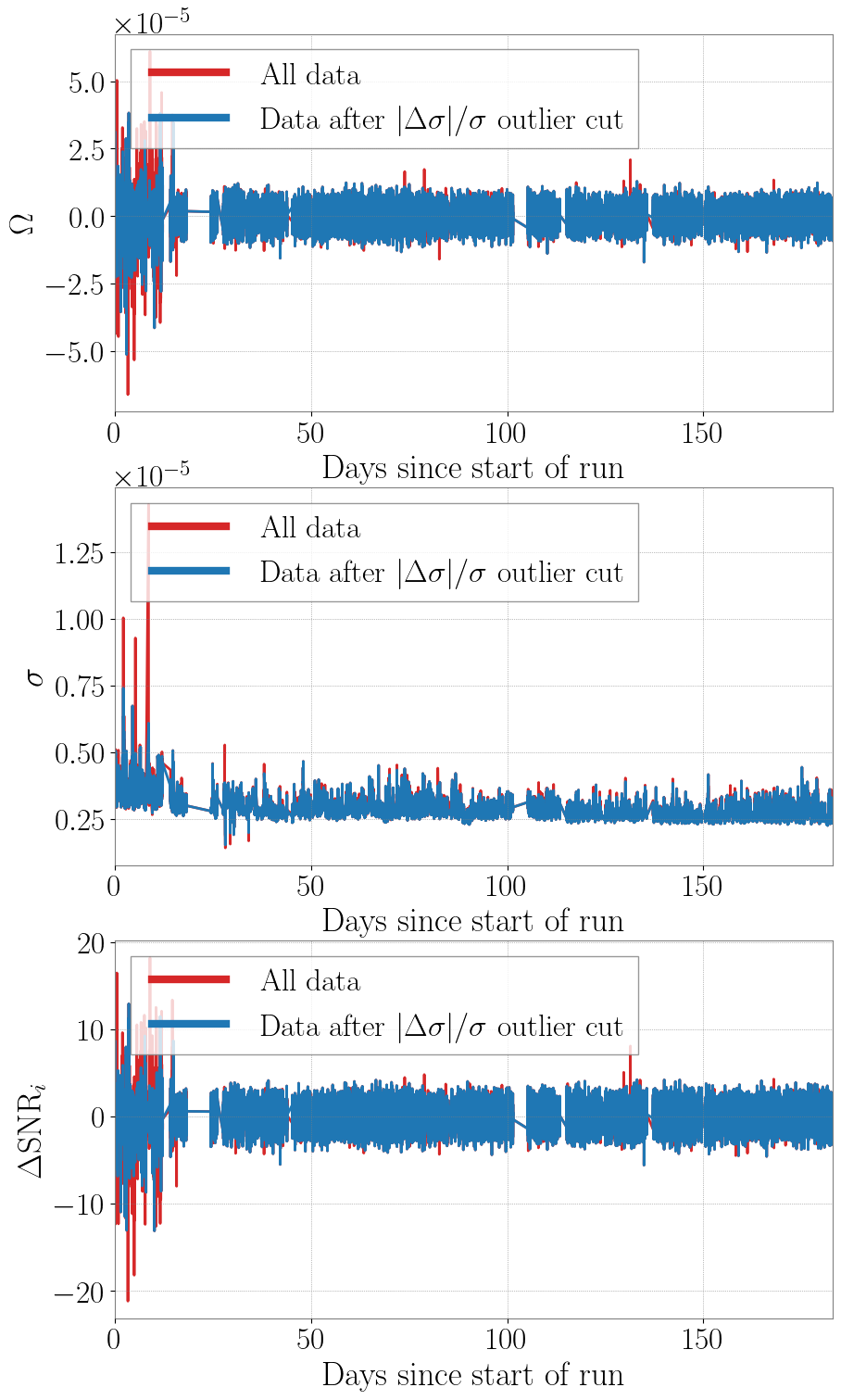}
    \raisebox{0.575\height}{\includegraphics[width=.43\textwidth]{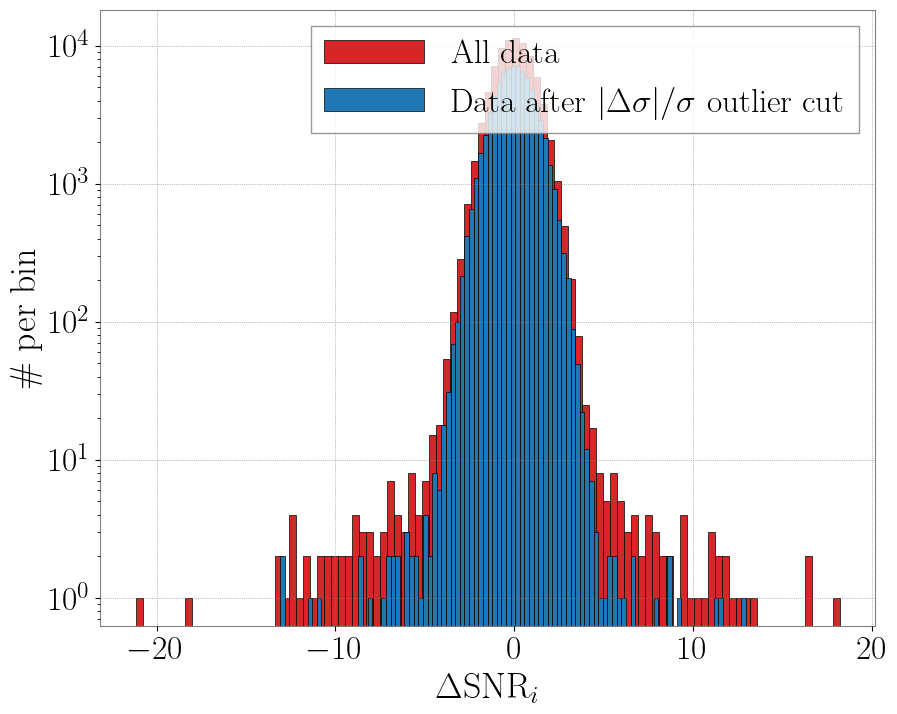}}
    \caption{\textbf{Left}: Point estimate, sigma and deviates $\Delta {\rm SNR}_i$ as a function of time before the delta-sigma cut (red) and after the cut (blue). \textbf{Right}: Distribution of the deviates $\Delta {\rm SNR}_i$ as a function of time before the delta-sigma cut (red) and after the cut (blue).}
    \label{fig:DscExamplePlots}
\end{figure}
The third set of checks produces the \gls{ift} of the point estimate spectrum, which should peak around zero seconds in case of a detection. 
Time-shifting the data in two detectors by more than the coherence time between the two detectors breaks the coherence between the two data streams, removing any evidence of a GWB signal. Note that the coherence time is determined by the bandwidth of our signal, which is of order 100 Hz, resulting in a coherence time of 10 ms. %
Hence, a \gls{gwb} signal will only peak around zero time lag between the output of the detectors. %

The fourth set studies the effect of the $\Delta\sigma$ data quality cut described in Sec. \ref{Sec:DeltaSigma} on the analysis run. %
To this end, we display several quantities before and after the cut is applied to the data, including the segment values of $\hat\Omega^\alpha_{{\rm ref}, i}$, $\sigma^\alpha_{{\rm ref}, i}$, and $\Delta\sigma^\alpha_{{\rm ref}, i}$, and the deviations in SNR,
\begin{equation}
    \Delta {\rm SNR}_i = \frac{\hat\Omega^\alpha_{{\rm ref}, i}-\langle\hat\Omega^\alpha_{{\rm ref}}\rangle}{\sigma^\alpha_{{\rm ref}, i}} \,,  
\end{equation}
as a function of time. %
Here angle brackets indicate an arithmetic mean over all segments $i$. %
We also plot a histogram of the values of $\Delta {\rm SNR}$ before and after the cut. %
This distribution should be centred around $0$, with a smooth narrower distribution after the application of the $\Delta\sigma$ cut. %
We additionally plot the $\Delta {\rm SNR}$ as a function of individual $\sigma^\alpha_{{\rm ref}, i}$. %
Finally, we plot the distribution of the ratios $\sigma^2_{{\rm ref}, i}/\langle\sigma^2_{{\rm ref}, i}\rangle$, which should peak around $1$. %
Some representative plots are shown as an example in Fig.~\ref{fig:DscExamplePlots}.

The last set of checks concerns a \gls{ks} test that is used to verify that the $\Delta {\rm SNR}_i$ are consistent with a Gaussian distribution. %
The \gls{ks} test implementation of this module returns the \gls{ks} test statistic, which is the maximal deviation from the Gaussian cumulative distribution function, as well as the p-value. %
These values can be used to make statements about the Gaussianity of the data \cite{KStestRef}. In addition, the cumulative distribution function is plotted for the data as well as for a Gaussian distribution. %

\section{Testing}

To comprehensively test the {\tt pygwb} analysis suite, we employ an efficient workflow to analyze datasets of increasing complexity. %
The datasets considered in this paper are:
\begin{enumerate}
    \item {\bf Continuous \gls{sgwb}:} A loud stationary and continuous stochastic signal generated with the {\tt simulator} module, injected in Advanced \gls{ligo} Hanford and \gls{ligo} Livingston assuming design A+ sensitivity~\cite{aplus_design_curve}. 
    \item {\bf Realistic \gls{cbc} \gls{gwb}:} A realistic background of merging \glspl{bbh} and \glspl{bns}, injected in Advanced \gls{ligo} Hanford and \gls{ligo} Livingston assuming design A+ sensitivity.
    \item {\bf O3 dataset:} The full Advanced \gls{ligo} Hanford and \gls{ligo} Livingston dataset from the third \gls{lvk} observing run~\cite{LIGO_O3}.
\end{enumerate}
The continuous \gls{sgwb} (dataset 1) is an idealized observing scenario, as our stochastic model matches the target signal perfectly by design, and as the signal is stationary and continuous our approach is optimal~\cite{Drasco-Flanagan:2003, Lawrence:2023buo}. %
The \gls{cbc} background (dataset 2) is a realistic scenario where the target signal is generated according to astrophysical models, informed by GW detections. %
In this case the signal is non-Gaussian, and we expect our approach to be un-biased~\cite{HLV-MDC-PhysRevD.92.063002, ET-first-MDC-PhysRevD.86.122001, ET-second-MDC-PhysRevD.89.084046} but sub-optimal~\cite{Drasco-Flanagan:2003, Lawrence:2023buo}, due to the intermittent nature of the signal which is not taken into account in the search method. %
For more details on the time-domain characteristics of these two types of signals and the detection challenges these present, see for example~\cite{Regimbau2022}. %
Finally, the O3 Advanced LIGO dataset (dataset 3) presents all the complexity of analyzing real GW detector data, which includes non-stationary noise, a large data volume, and expensive computational requirements. %

We handle large datasets by splitting the data into smaller {\tt pygwb$\_$pipe} jobs, assuming each job is independent; %
these are then combined using the {\tt pygwb$\_$combine} script (see Sec.~\ref{sec: pipeline} for details). %
We then employ a parameter estimation script, {\tt pygwb$\_$pe}, based on the {\tt pe} module described in Sec.~\ref{sec:pe}, to perform parameter estimation on specific models. %
For more details on how to run this sort of analysis, we refer users to the online documentation for the most up-to-date workflow instructions~\cite{docs}. %
In the following, we present the different datasets and summarize our analysis results.

\subsection{Mock data}
\label{sec: MDC}

\subsubsection{Stationary and continuous stochastic gravitational-wave background}
\label{MDCSimulator}
We employ the {\tt Network} (Sec.~\ref{sec:network}) to generate a stationary and continuous \gls{sgwb} signal with a fixed \gls{psd}, $S_h(f)$. %
This allows us to simultaneously test the module and the whole analysis pipeline. %
The injected \gls{sgwb} is scale-invariant, i.e., $\Omega_{\rm GW}(f)$ is constant over frequencies, 
\begin{equation}
    \Omega_{\rm inj}(f)= 1.06 \times 10^{-7}\,.   
\end{equation}
This is converted to $S_h(f)$ using the relation in Eq.~\eqref{eq:omegatoI}. %
The noise $P_n(f)$ is taken to be Gaussian, colored using the the Advanced \gls{ligo} noise \gls{psd} \cite{aLIGO_sensitivity}. %
One hundred days of consecutive data are simulated at a sampling rate of $1024$ Hz. %

\begin{figure}[t]
    \centering
    \includegraphics[width=0.45\textwidth]{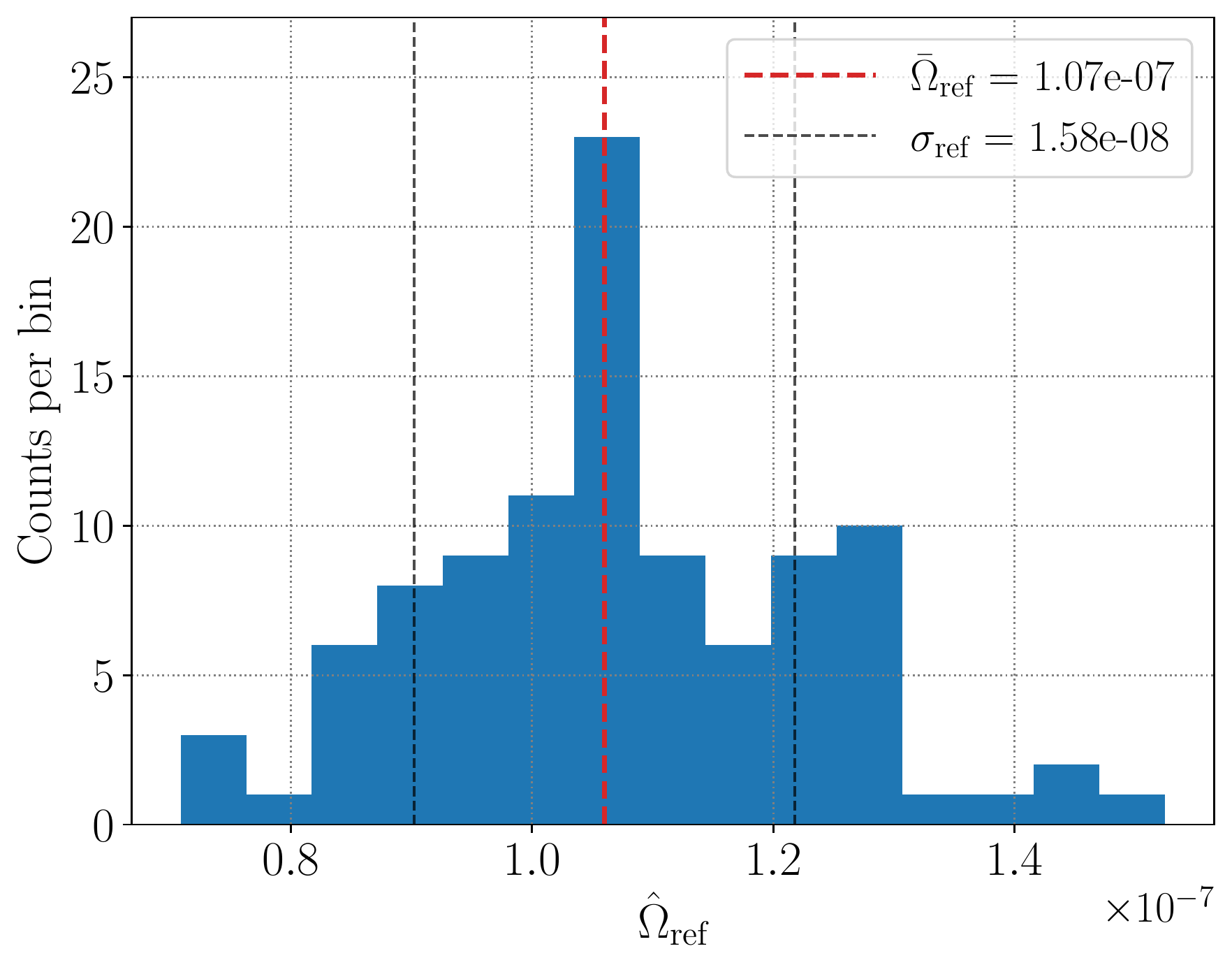}
    \includegraphics[width=0.3\textwidth]{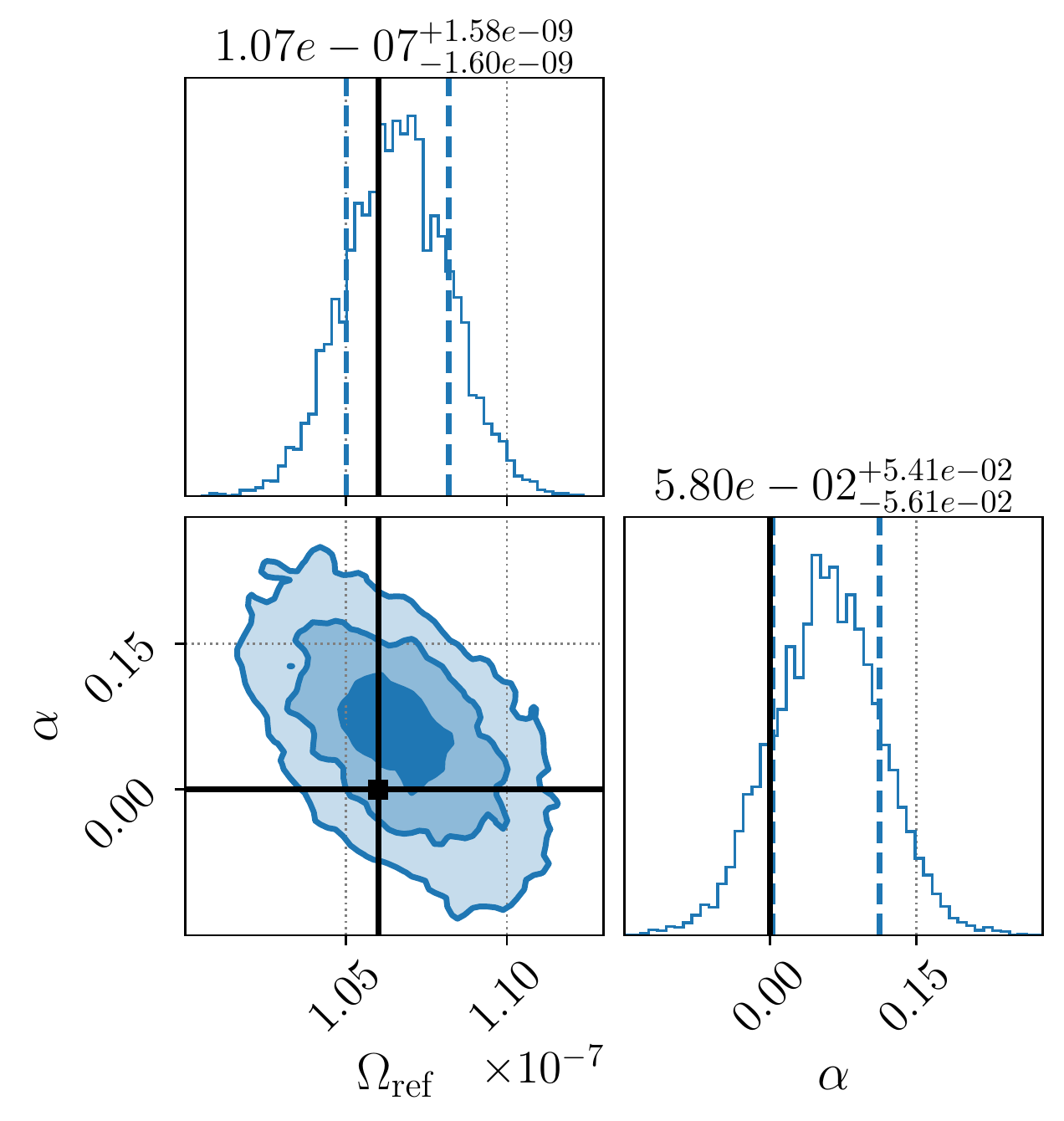}
    \caption{\textbf{Left}: Distribution of the recovered point estimate for each day in the dataset. The injected value is denoted by the red line. $\overline{\Omega}_{\rm ref}$ and $\sigma_{\rm ref}$ denote the mean and the standard deviation of the one hundred point estimates. \textbf{Right}: Parameter estimation performed on the one hundred days, obtained assuming a log-uniform prior from $10^{-11} - 10^{-6}$ for $\Omega_{\rm ref}$ and Gaussian prior with mean 0 and standard deviation 1.5 for $\alpha$. The injected values are denoted by the black lines, while the contours represent the 1$\sigma$, 2$\sigma$, and $3\sigma$ contours. The vertical dashed lines represent the $1 \sigma$ confidence interval.}
    \label{fig:hundreddaydataset}
\end{figure}

Each of the one hundred days is analyzed separately, and we recover a distribution of $\hat\Omega^0_{\rm 25}$ point estimates, shown in Fig.~\ref{fig:hundreddaydataset} (left), using $\alpha=0$ and $f_{\rm ref} = 25$ Hz in the pipeline. %
Analyzing one hundred days separately allows us to construct a distribution of recovered point estimates, which is useful to assess the ability of the {\tt simulator} module to inject a stochastic stationary signal. %
We then perform parameter estimation on the combined one hundred days, presented in Fig. \ref{fig:hundreddaydataset} (right). We assume a log-uniform prior from $10^{-11} - 10^{-6}$ for $\Omega_{\rm ref}$ and Gaussian prior with mean 0 and standard deviation 1.5 for $\alpha$. %
This shows a recovery within $1\sigma$ for $\Omega_{\rm ref} = 1.06 \times 10^{-7}$ and within $2\sigma$ for the spectral index $\alpha_{\rm inj} = 0$. 

The tests above illustrate that the {\tt simulator} module is able to successfully inject a stochastic stationary signal and that the {\tt pygwb} pipeline is able to recover this injection.

\subsubsection{Gravitational-wave background from a coalescing compact binary population}

The inspiral and merger of two compact objects emit a characteristic \gls{gw} signal. %
We generate datasets containing a \gls{gwb} signal resulting from the superposition of \gls{gw} signals from a set of \gls{cbc} populations including \glspl{bbh} and \glspl{bns}. %
 To simulate the signals, we employ the code used in the past for the \gls{et} \glspl{mdsc} \citep{ET-first-MDC-PhysRevD.86.122001, ET-second-MDC-PhysRevD.89.084046, ET-second-MDC-low-f-PhysRevD.93.024018} and for the Advanced \gls{ligo} and Advanced Virgo \gls{mdsc} \citep{HLV-MDC-PhysRevD.92.063002}. %
The Monte Carlo algorithm that we use for the generation of a compact binary population up to redshift $z=10$ is extensively described in \cite{ET-first-MDC-PhysRevD.86.122001} and \cite{Regimbau:2014nxa}. We summarize below the main steps of the simulations.

To generate a \gls{cbc} population we assume a merger rate per unit redshift~\citep{Belczynski:2006br, Berger:2006ik, Belczynski:2000wr, Bulik:2003kr},
\begin{equation}
    \frac{\text{d}R(z)}{\text{d}z} = \frac{\text{d}V_{\rm c}}{\text{d}z} {r_c}(z),
\end{equation}
where $\text{d}V_{\rm c}/\text{d}z$ is the co-moving volume element and ${r_c}$ the coalescence rate as a function of redshift~\cite{Regimbau:2011rp}. %
The element of co-moving volume assumes a $\Lambda$CDM cosmology from Planck 2018~\cite{Planck2018} (Hubble parameter $H_0 = 67.7\, \mathrm{km/s/Mpc}$, $\Omega_{m}=0.31$ and $\Omega_{\Lambda}=1-\Omega_{m}$). %
We assume a coalescence rate normalized to a local rate $r_c(0)=1\,\mathrm{Mpc^{-3}\, Myr^{-1}}$ for \gls{bns} coalescences and  $r_c(0)=3\,\mathrm{Mpc^{-3}\, Myr^{-1}}$ for \gls{bbh} coalescences, assuming the star formation rate from \cite{Hopkins:2006bw} and a minimum delay time between binary formation and merger of $20\, \mathrm{Myr}$ for \glspl{bns} and $50\, \mathrm{Myr}$ for \glspl{bbh}; see \citep{Dominik:2012kk, Neijssel_2019} for more details. %
These choices give rise to a dataset composed by $87 \%$ of \glspl{bns} and $13 \%$ of \glspl{bbh}. %

The time intervals $\tau$ between consecutive \gls{cbc} events in our population are obtained by sampling an exponential distribution $P(\tau) = \exp(-\tau/\bar{\tau})$, where $\bar{\tau}$ is the average time between consecutive events. This is consistent with the assumption that the coalescence times $t_c$ of the events behave as a Poisson process~\cite{Regimbau:2014nxa}. %
The coalescence redshift is drawn from the normalized coalescence rate $p(z)=\bar{\tau}\text{d}R/\text{d}z(z)$ within $z \in [0,10]$. %
The sky position $\hat{n}$ of each source is generated isotropically on the sky. %
The \gls{gw} polarization angle $\psi$, the phase angle $\phi_0$ at the coalescence time, and the cosine of the inclination angle of the orbital plane to the line of sight $\iota$ are all drawn from uniform distributions. %
The mass function of the components in the \glspl{bbh} is chosen to be a \gls{plpp} from the preferred case presented in the \gls{lvk} collaboration \gls{cbc} population inference paper~\citep{O3_pop_paper_LIGOScientific:2021psn} or a simple \gls{pl}~\citep{O2_pop_paper_LIGOScientific:2020kqk}, while the \gls{bns} masses are drawn from uniform distribution between 1 and 3 $M_{\odot}$. 
The \gls{bbh} mass functions are used to label the two datasets presented below. %
Spins are neglected in both cases. %
\begin{figure}[t]
    \centering
\includegraphics[width=0.5\textwidth]{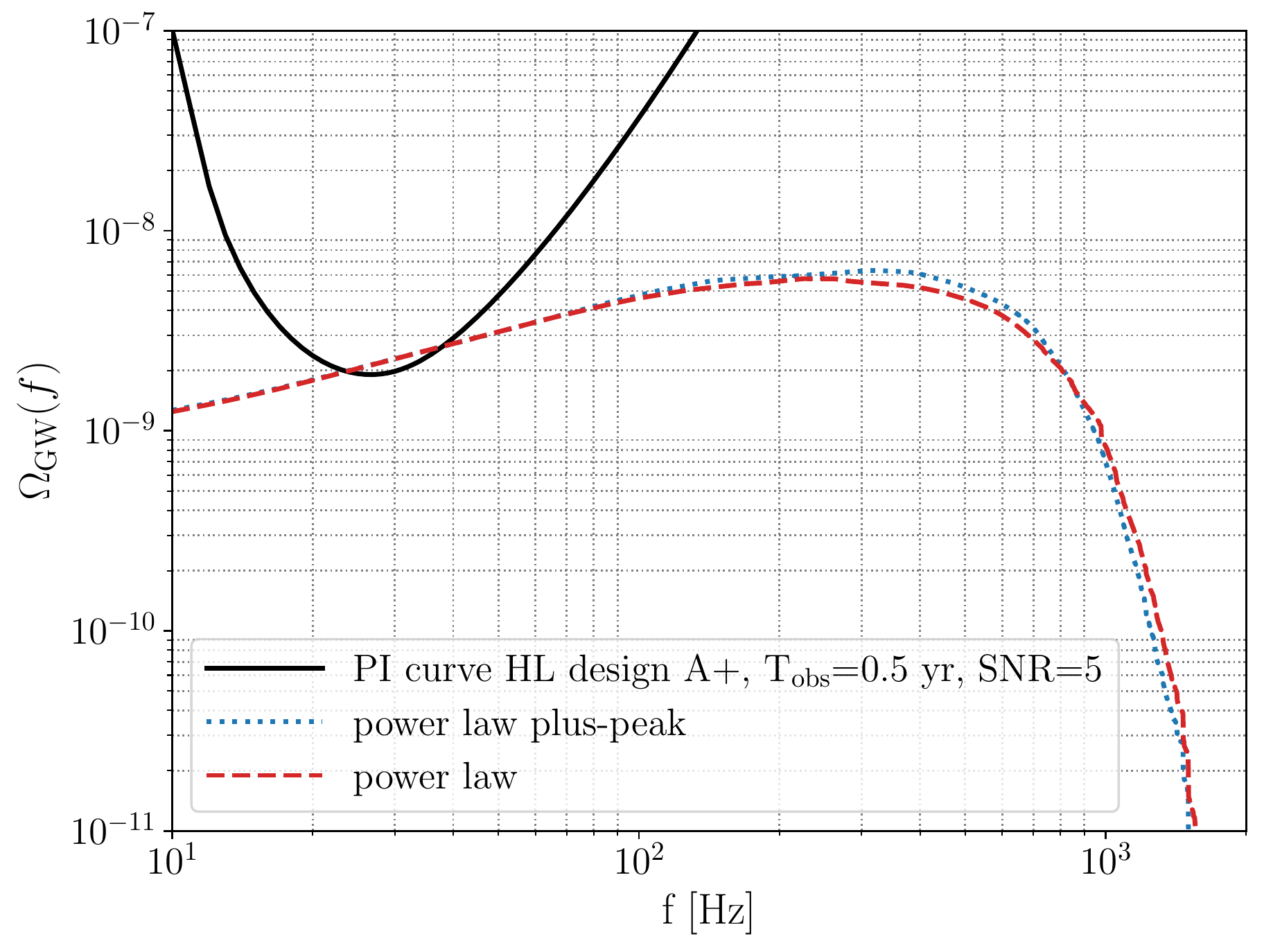}
    \caption{$\Omega_{\mathrm{GW}}(f)$ for the dataset corresponding to the \gls{plpp} (blue) and the \gls{pl} (red) models for the mass function. The black line is the power-law integrated sensitivity (PI) curve for an observation time of six months and an expected \gls{snr} = 5, assuming the HL baseline with Advanced \gls{ligo} plus design sensitivity. The simulated signals intersect the PI curve, hence they are expected to be detected with an \gls{snr} of at least 5.}
    \label{fig:Omega_CBC_injection}
\end{figure}

For each source, the signal waveform is generated in the time domain. %
For \glspl{bns}, we use the {\tt TaylorT4} time-domain waveform \citep{TaylorT4_Buonanno:2002fy}. %
For \gls{bbh} signals, we use the {\tt EOBNRv2} \citep{EOBNRv2_Buonanno:2009zt} %
time-domain waveform from numerical relativity. %
These are then injected into the \gls{ligo} Hanford and Livingston detectors, with the addition of colored Gaussian noise generated from the \gls{ligo} A+ Design~\cite{MDC_sensitivity_curves, aplus_design_curve} expected sensitivity curve, to produce the final datasets. %

Following the above prescriptions, we generate two six-months datasets with sampling frequency 1024 Hz, labelled \gls{plpp} and \gls{pl}, formed by two different \gls{cbc} populations. %
The two populations differ by the mass distributions of the \glspl{bbh} and the average time between consecutive events, as seen in Table \ref{tab:MDC_CBC_results}. %
The latter is chosen such that the \gls{gwb} amplitudes of the two datasets match, for ease of comparison. %
Furthermore, to obtain a \gls{snr} large enough to confidently detect the injected \gls{gwb}, a small amplification of the signal is required. %
To this end, the amplitude of the \gls{cbc} waveforms is multiplied by 1.5 and 1.7 for the \gls{plpp} and the \gls{pl} datasets, respectively, resulting in an injected value of $\Omega_{\mathrm{ref}} = 2.05 \times 10^{-9}$. %

The $\Omega_{\mathrm{GW}}(f)$ spectrum relative to the each dataset is obtained by summing the contributions from individual coalescences \citep{HLV-MDC-PhysRevD.92.063002}, and is illustrated in Fig.~\ref{fig:Omega_CBC_injection}. %
As may be observed, in the case of \gls{cbc} signals $\Omega_{\mathrm{GW}}(f)$ increases as $f^{2/3}$ from the inspiral phase (and then as $f^{5/3}$ from the \gls{bbh} merger phase) before reaching a peak and steeply decreasing~\cite{Marassi:2011si}. %
This motivates fixing the spectral index parameter to $\alpha=2/3$ in our searches. %
Fig.~\ref{fig:Omega_CBC_injection} also shows the \gls{pi} curve \citep{PI_curves:PhysRevD.88.124032} for the Hanford-Livingston baseline, assuming the design A+ sensitivity for the two detectors \citep{aplus_design_curve}, an observation time $T_{\rm obs} = 6$ months, and a desired sensitivity of \gls{snr}=5. %
Given that the \gls{pi} curve is almost tangent to $\Omega_{\mathrm{ref}}$ of the two datasets, we expect to observe the \gls{gwb} signals with \gls{snr} $\sim 5$. %


\begin{figure}[t]
    \centering
    \includegraphics[width=0.4\linewidth]{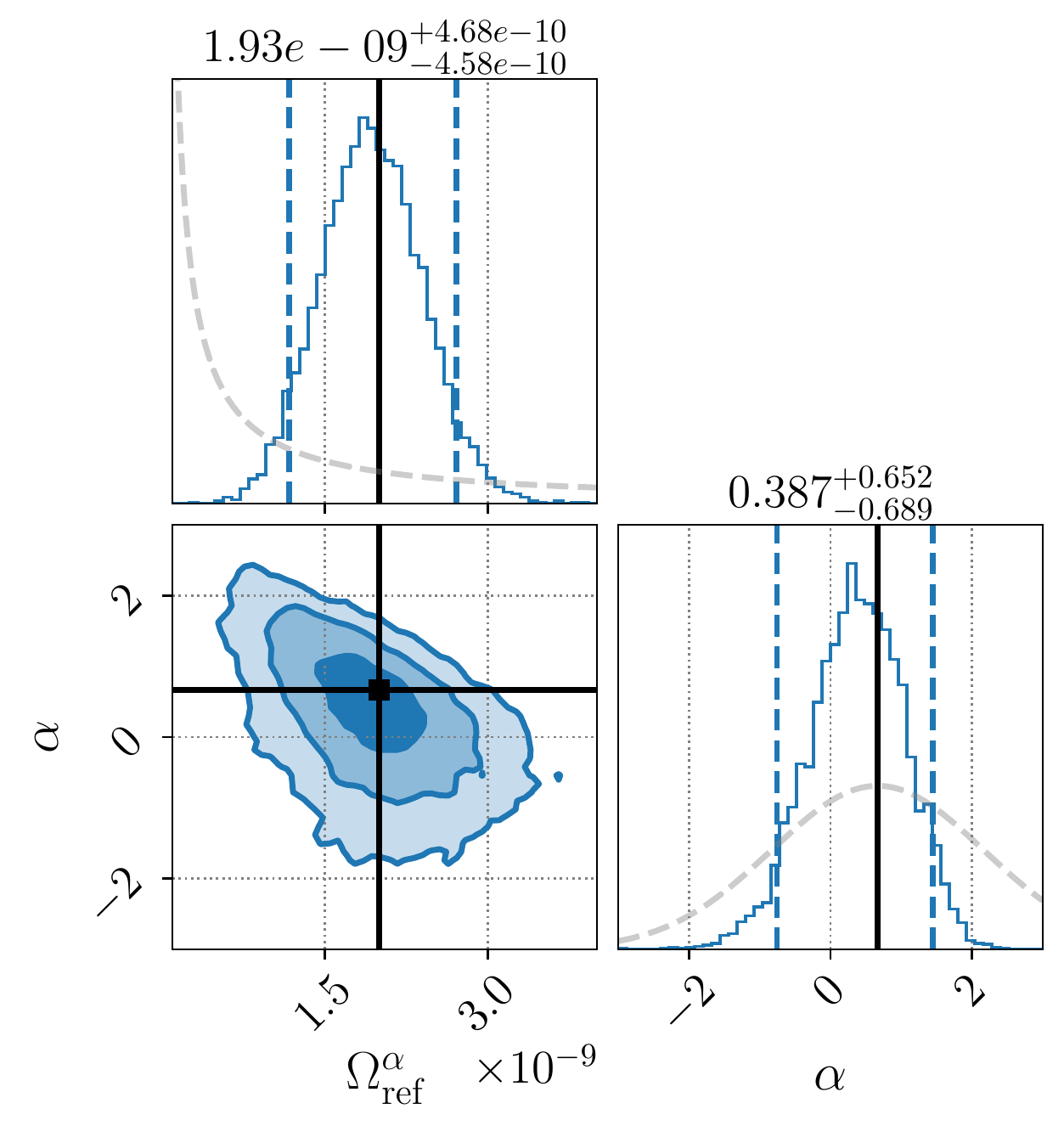}
    \includegraphics[width=0.4\linewidth]{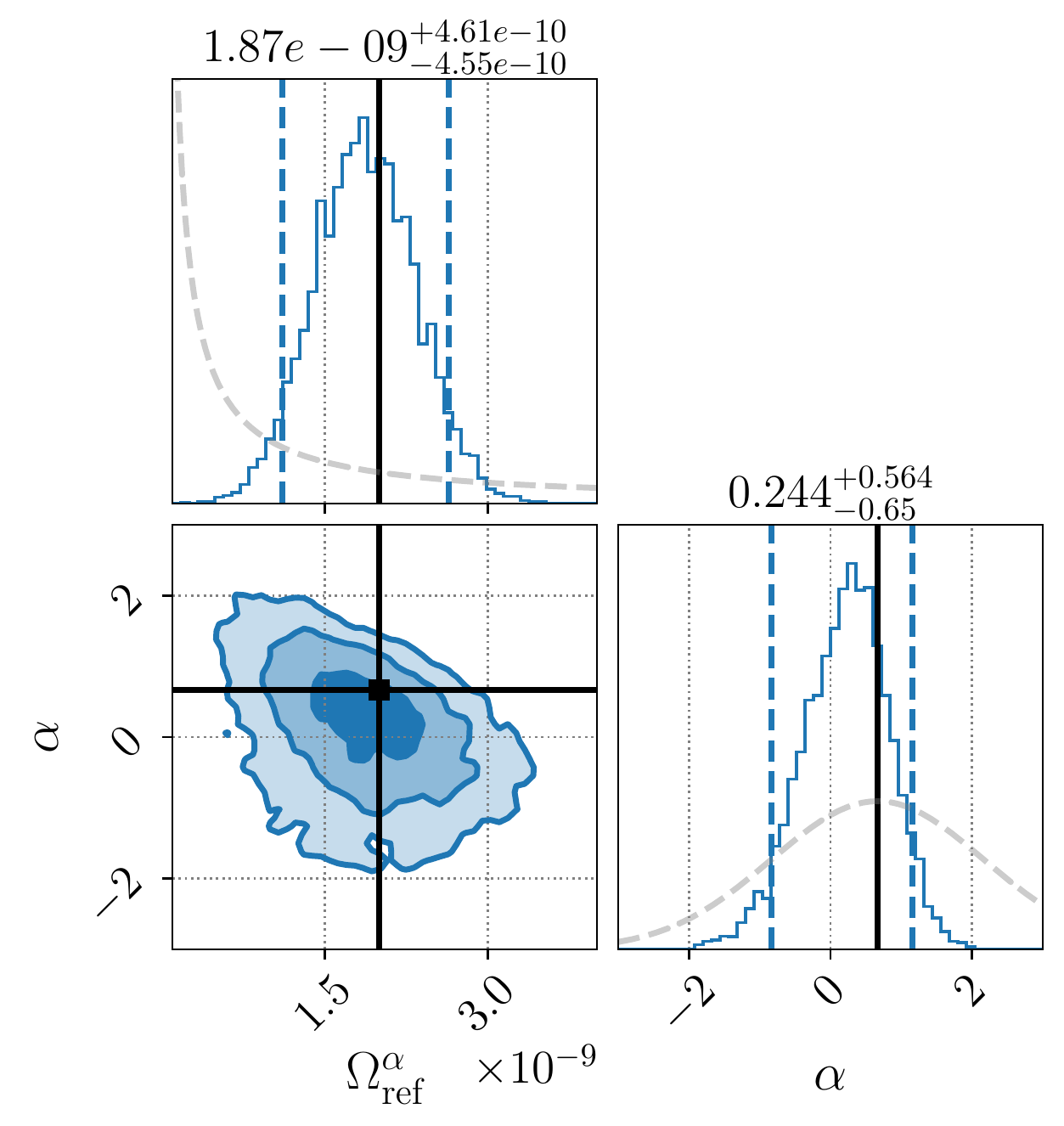}
    \caption{PE results. Left: Corner plot obtained from running the parameter estimation over the \gls{plpp} dataset.
    Right: Corner plot obtained from running the parameter estimation over the \gls{pl} dataset.
    Each plot shows the posteriors on $\Omega_{\mathrm{ref}}$ and $\alpha$ obtained assuming a log-uniform prior on $\Omega^0_{\mathrm{ref}}$ from $10^{-11}$--$10^{-8}$ and a Gaussian prior on $\alpha$ with mean 2/3 and standard deviation of 1.5, respectively, denoted by the gray dashed lines. The injected values are represented by the black lines, indicating a recovery of both the amplitude of the signal and $\alpha$ within $1\, \sigma$. The vertical blue dashed lines represent the $2 \sigma$ confidence interval.}
    \label{fig:MDC_pe}
\end{figure}
We analyze the two datasets in the frequency band $20 - 500$ Hz, using a frequency resolution of $1/32$ Hz and a segment duration of 192 s. %
We choose $\alpha = 2/3$, $f_{\rm ref} = 25$ Hz, and $H0=67.7$ km/s/Mpc for this analysis. %
The results of the analysis are summarized in Table \ref{tab:MDC_CBC_results}. %
We recover the \gls{plpp} injection within $1\, \sigma$, and observe it with \gls{snr} = 5.4, while recovering the \gls{pl} injection within $1\, \sigma$, with \gls{snr} = 5.0. %
We attribute the differences in the recoveries to the specific data and noise realizations within the datasets. %

We then proceed with estimating the parameters $\alpha$ and $\Omega^\alpha_{\mathrm{ref}}$ modelling $\Omega_{\mathrm{GW}}(f)$ as a simple power-law in frequency as given by Eq.~\eqref{eq:omega-plaw}. %
We assume a log-uniform prior over $\Omega^0_{\mathrm{ref}}$ in the range $[\Omega^0_{\mathrm{min}},\,\Omega^0_{\mathrm{max}}]=[10^{-11}, \,10^{-8}]$, and a Gaussian prior on $\alpha$ with mean $2/3$ and standard deviation $(\log_{10}{\Omega^0_{\mathrm{min}}}-\log_{10}{\Omega^0_{\mathrm{min}}})/2 = 1.5$. %
Note that the priors in $\Omega^\alpha_{\mathrm{ref}}$ are defined for $\alpha = 0$. %
The choice of the prior over $\alpha$ can be understood as follows. %
The log-uniform prior over $\Omega^0_{\mathrm{ref}}$ induces some implicit prior over $\alpha$ that can be shown to be a triangular prior centred on $\alpha = 0$ and non-zero for $|\alpha| \leq (\log_{10}{\Omega^0_{\mathrm{max}}}-\log_{10}{\Omega^0_{\mathrm{min}}})$. %
To avoid a vanishing prior outside of this range, we choose a Gaussian prior for $\alpha$ with standard deviation comparable with the triangular prior, centered on $\alpha =2/3$ to better match the injected \gls{gwb}. %

Parameter estimation corner plots are shown in Fig.~\ref{fig:MDC_pe}. %
For both datasets, $\Omega_{\mathrm{ref}}$ and $\alpha$ are recovered within $1\, \sigma$. %
The log-Bayes factors $\mathcal{B}^{\rm GW}_{\rm noise}$ are $11.1$ and $9.2$ for the \gls{plpp} and \gls{pl} datasets, respectively, indicating strong evidence~\cite{bayes_factor_doi:10.1080/01621459.1995.10476572} for the presence of signal over noise only. 

\begin{table}[h]
    \centering
    \begin{tabular}{c|c|c|c|c|c}
       {\sc dataset}  & $\tau$ (s) & $a$ &  $  ~\qty(\hat{\Omega}^{2/3}_{25} \pm \hat{\sigma}^{2/3}_{25}) \times 10^{9}$~ & ~SNR~ & ~$\mathcal{B}^{\mathrm{GW}}_{\mathrm{noise}}$~\\
    \hline
    \hline
        PLPP & ~60~  & ~1.5~  & $2.09 \pm 0.39$ & ~5.4~ & ~11.1~\\
        PL & ~54.7~ & ~1.7~ & $1.94 \pm 0.39$ & ~5.0~ & ~9.2~
    \end{tabular}
    \caption{Parameters and results of each dataset. The first row refers to the \gls{plpp} dataset, while the second row to the \gls{pl} one. The second and third columns display the average time between two successive binary mergers, $\tau$, and the waveform amplification factor, $a$. The last three columns illustrate the recovered point estimate with $1\, \sigma$ uncertainty on the quantity $\hat{\Omega}_{\mathrm{ref}}^\alpha$ ($f_{\mathrm{ref}} = 25\, \mathrm{Hz}$, $\alpha=2/3$), the corresponding \gls{snr}, and the log-Bayes factor $\mathcal{B}^{\mathrm{GW}}_{\mathrm{noise}}$.
    }
    \label{tab:MDC_CBC_results}
\end{table}
\subsection{O3}
\label{Sec:O3Data}

In this section we present results from the application of the {\tt pygwb} analysis suite to the full \gls{ligo} Hanford and \gls{ligo} Livingston O3 dataset~\cite{gwosc}. %
We set upper limits on the signal from a \gls{sgwb} and confirm these are consistent with previously published collaboration results~\cite{LIGO_O3}.

The O3 data run collected between April 1, 2019 and March 27, 2020, divided into two sub-sets with an interruption between October 1 and November 1, 2019, with a total coincident livetime of 205.4 days between \gls{ligo} Hanford and \gls{ligo} Livingston. %
These are reduced to 196 days after {\it category 1 vetoes}\footnote{``Category 1'' vetoes flag data which are unsuitable for analysis, such as incorrectly calibrated data, data collected during atypical operation of the instruments, and data with severe data quality issues.}~\cite{Virgo:2022kwz, Abbott_2018} and external non-stationarity cuts are applied (for details, see~\cite{LIGO_O3}). %
The {\tt pygwb} analysis is implemented with the workflow described above. %
The O3 data, natively sampled at 16384 Hz, are downsampled to 4096 Hz and high-pass filtered at 11 Hz. %

\begin{figure}[t]
    \centering
    \includegraphics[width=0.75\textwidth]{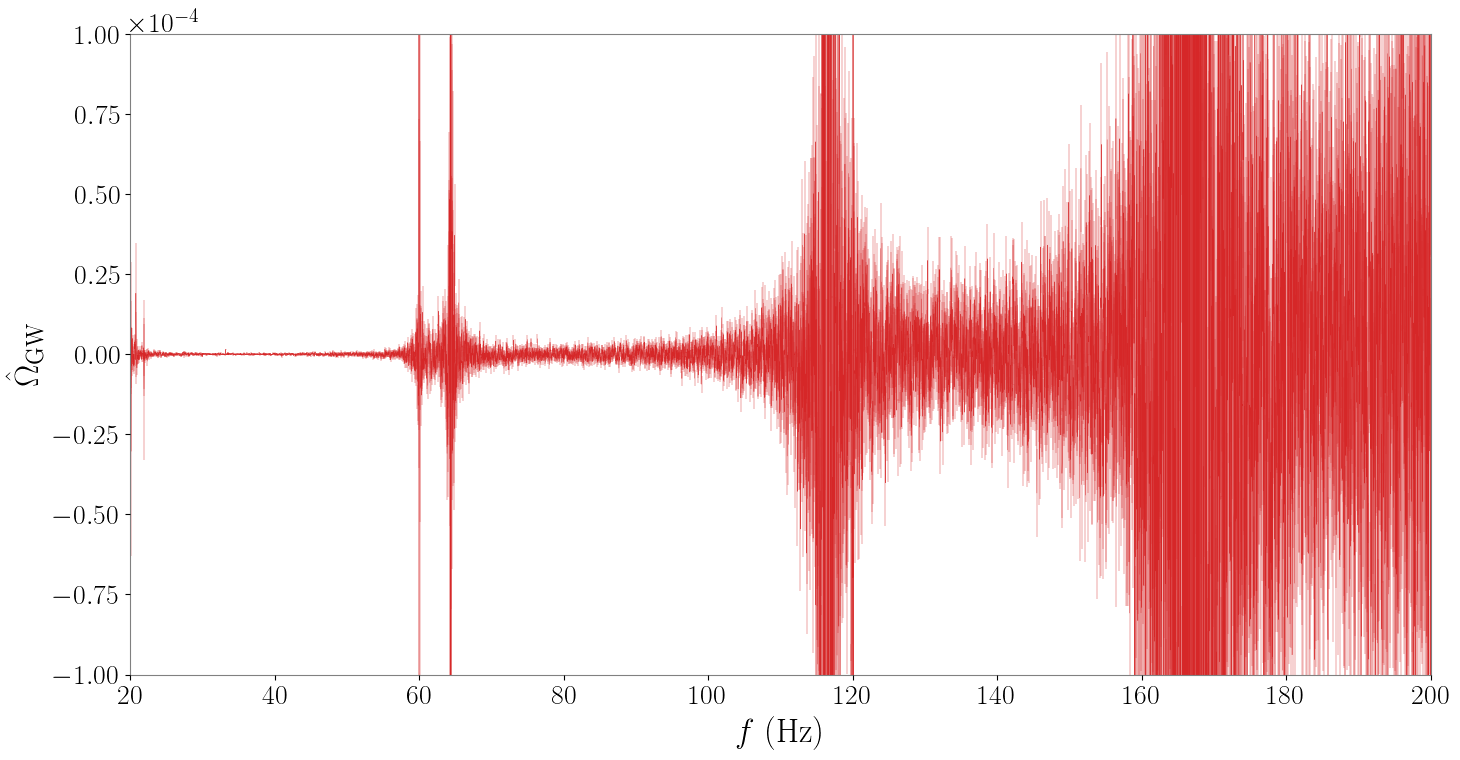}
    \caption{Estimated cross-correlation spectrum $\hat{\Omega}^0_{25}\pm \hat{\sigma}^0_{25}$ from O3 data. By eye, it is possible to spot several narrowband artifacts (lines) which are subsequently excluded from our analysis.}
    \label{fig:O3_crosscorr_spectrum}
\end{figure}

The time-averaged O3 \gls{ligo} Hanford -- Livingston cross-correlation spectrum is presented in Fig.~\ref{fig:O3_crosscorr_spectrum}. %
Our $\Delta\sigma$ threshold excludes 7.8\% of the analyzed time (see Sec.~\ref{Sec:DeltaSigma} for implementation details). %
This result matches the previous stochastic non-stationarity cut published in~\cite{LIGO_O3} within 1\%, with the previous cut excluding an extra 0.06\% of the time. %
We believe this small variation to be due to a different window bias factor used in the two analyses (the bias factor calculation used here is outlined in App.~\ref{sec:app_window}). %

We calculate broad-band integrated estimates between $20 - 1726$ Hz of $\Omega_{\rm GW}(f_{\rm ref})$ for different power-law spectral models, applying the released O3 notchlist~\cite{O3IsotropicDataset} to exclude known problematic frequencies~\cite{Covas_2018}. %
A summary of the values for the point estimate and uncertainty for these is presented in Table~\ref{tab:O3_results}. %
The uncertainties $\sigma^\alpha_{\rm ref}$ agree within $1\%$ with previously published LVK results, presented in~\cite{LIGO_O3}. %
The point estimates for $\Omega^\alpha_{\rm ref}$ fluctuate notably more than the uncertainties. %
We believe this to be due to small differences in the analyses, to which the point estimates are more sensitive, such as individual start and end time of each pipeline job, and the differences in the non-stationarity cuts described above. %

\begin{figure}
    \centering
    \includegraphics[width=0.4\linewidth]{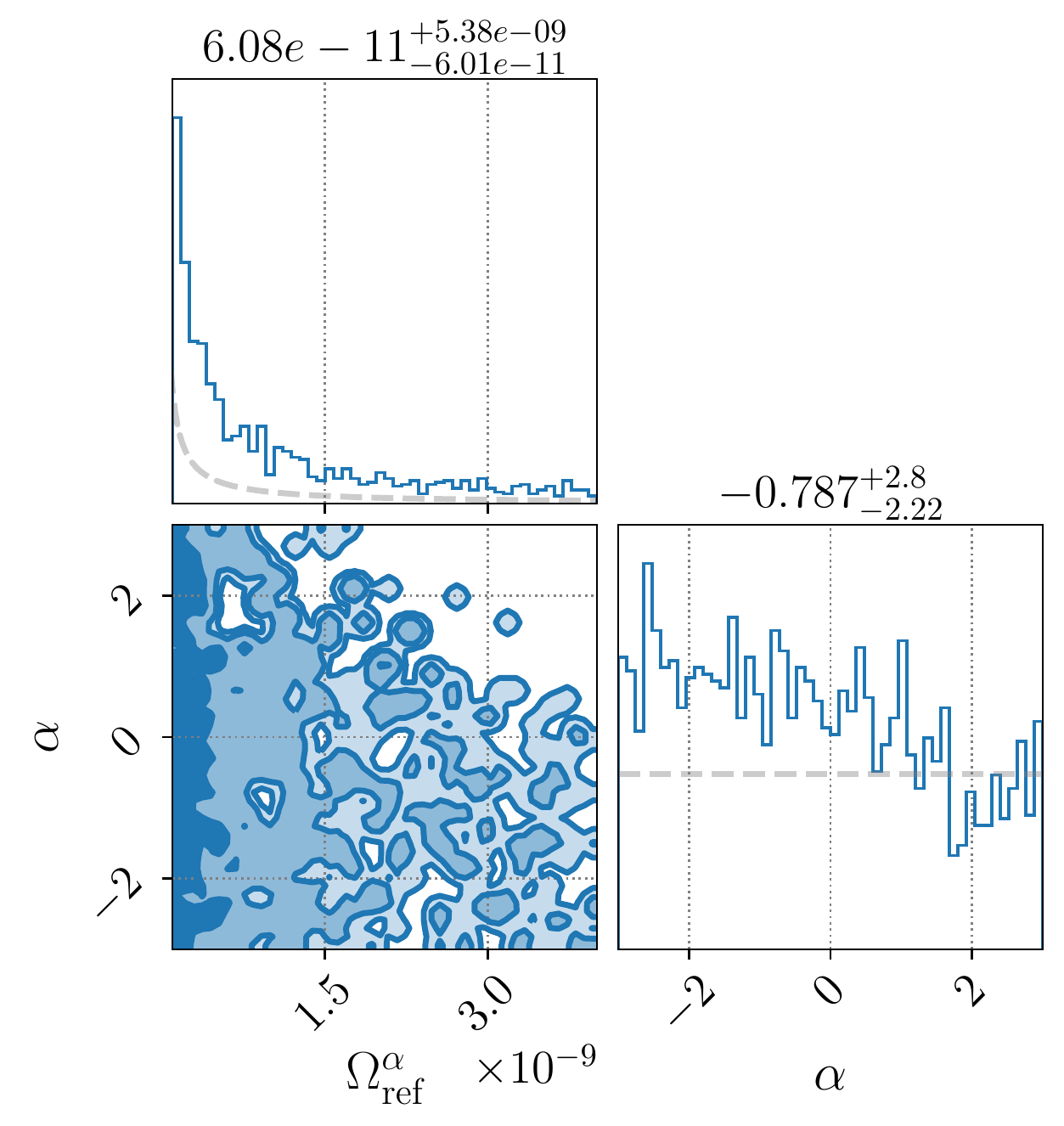}
    \includegraphics[width=0.4\linewidth]{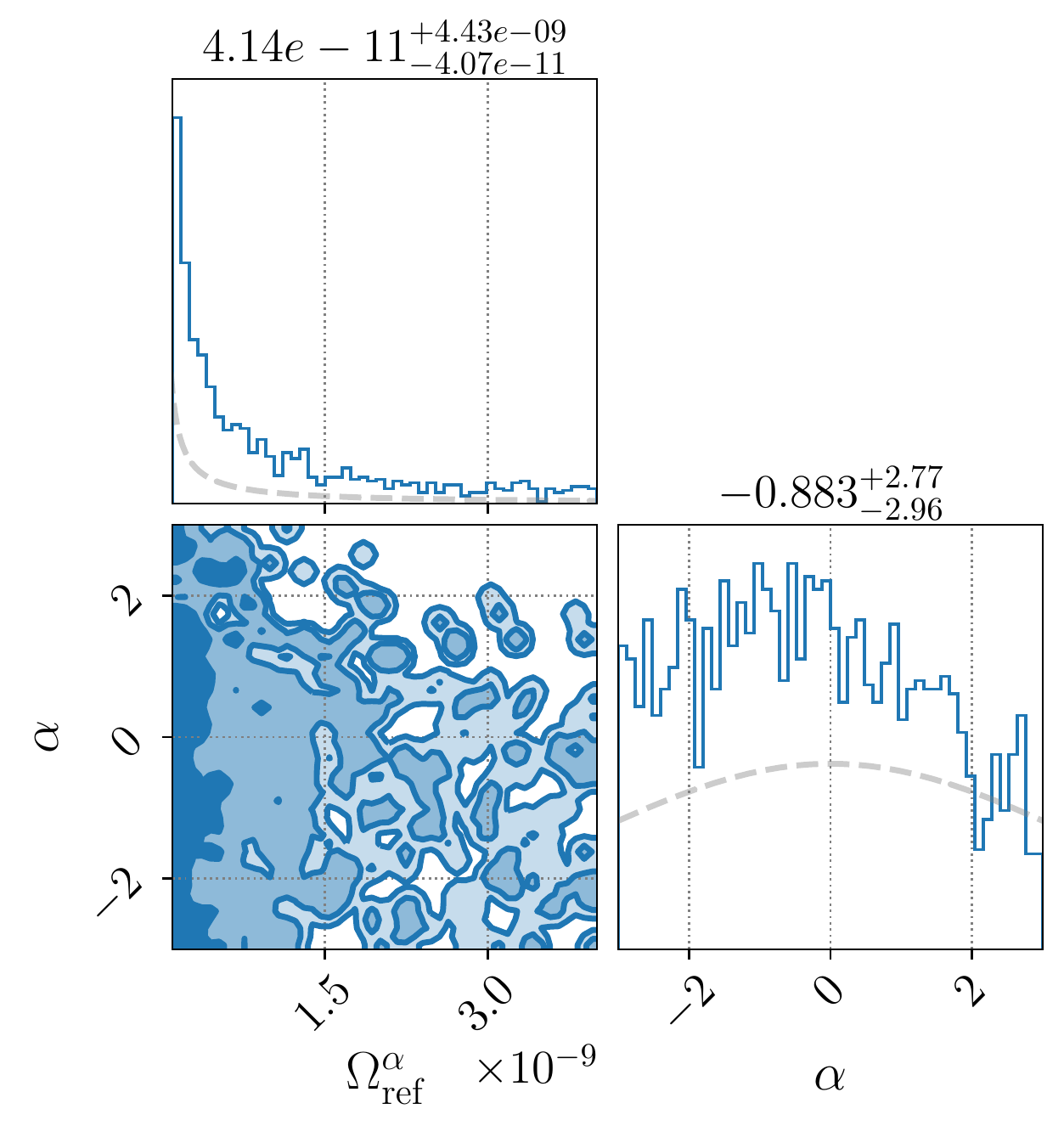}
    \caption{Parameter estimation results with {\tt pygwb$\_$pe} on LVK O3 data, using a log-uniform prior on $\Omega_{25}$, and a uniform prior (left) or a Gaussian prior on $\alpha$ (right), as described in the text. The priors are denoted by the gray dashed lines. }
    \label{fig:O3_pe}
\end{figure}

Finally, we perform parameter estimation to constrain $\Omega_{\rm GW} (f_{\rm ref}= 25 \text{Hz})\equiv \Omega_{25}$ and the spectral index $\alpha$ with O3 data. %
We employ a log-uniform prior on $\Omega_{25}$ spanning $[10^{-13},10^{-6}]$, and present results for two different priors on $\alpha$: a uniform prior between $[-4, 4]$ and a Gaussian prior centered around 0 with norm 3.5 (the latter matches the choice in ~\cite{LIGO_O3}). %
To account for calibration uncertainty, we marginalize over the uncertainty parameter $\lambda$ as described in App.~\ref{sec:app_calibration}, with combined calibration error for Hanford and Livingston of $1.48\%$, as in~\cite{LIGO_O3}. %
Parameter estimation confirms $\Omega_{25}$ is consistent with 0 and $\alpha$ remains unconstrained, as may be seen in Fig.~\ref{fig:O3_pe}. %
These results agree with the previous parameter estimation carried out in~\cite{LIGO_O3}.

Our results are quoted at the value of the Hubble parameter $H_0 = 67.9$ km/s/Mpc, in line with published results. %
This is not the built-in value of $H_0$, defined in Sec.~\ref{sec:postproc}; however, rescaling is straightforward as it is an overall multiplication factor, which may be changed when post-processing the run with the {\tt pygwb$\_$combine} script or manually using the built-in functions of the {\tt OmegaSpectrum} object, as explained in Sec.s~\ref{sec:postproc} and~\ref{sec: pipeline}.

We would like to note that this entire analysis was carried out on a large computing cluster and completed in less than five hours of human time. %
This is an example of the computational efficiency of our package.

\begin{table}[h]
    \centering
    \begin{tabular}{c|c|c||c|c|c}
       ~$\alpha$~  &  ~$\hat{\Omega}^\alpha_{25}\times 10^9$~ & ~$\hat{\Omega}_{\rm LVK}\times 10^9$ & ~prior~ & ~$\Omega_{\rm pe}$ ($95\%$ UL) ~& ~$\alpha_{\rm pe}$~\\
       \hline 
       \hline 
       
         0 & $-3.4\pm 8.1$& $-2.1\pm 8.2$ & {\sc uniform} & $5.44\times 10^9$ & $-0.8^{+2.8}_{-2.2}$\\
         2/3 & $-4.5\pm 6.1$& $-3.4\pm 6.1$ & ~{\sc gaussian}~ & $4.06\times 10^9$ & $-0.5^{+2.8}_{-2.8}$ \\
         3 & $-1.5\pm 0.9$& $-1.3\pm 0.9$ & & 
    \end{tabular}
    \caption{Summary of {\tt pygwb} search results on O3 dataset. On the left, three columns summarising point estimates from the weighted optimal statistic, at different spectral indices $\alpha$. On the right, three columns summarising Bayesian upper limits (UL) with log-uniform prior on $\Omega^0_{25}$ and either uniform or Gaussian prior on $\alpha$. These results are consistent with no detection of the amplitude of the background ($\Omega^\alpha_{\rm ref}$ is consistent with 0), nor its spectral shape ($\alpha$ remains unconstrained).}
    \label{tab:O3_results}
\end{table}

\section{Conclusions}
We present a new Python--based package tailored to \gls{gwb} searches with current ground-based interferometers. %
We opt for a modular code, where each module performs specific tasks of the \gls{gwb} data analysis. %
The modularity of the code results in large flexibility and offers the possibility to customize the pipeline according to one's own needs. %
With the use of Python language, the user-friendliness and flexibility of the code, we aim to bring \gls{gwb} searches to the wider \gls{gw} community, as the detection of a \gls{gwb} with ground-based interferometers draws potentially closer. %
With increasing amounts of \gls{gw} data, {\tt pygwb} also answers the need for an open-source \gls{gwb} data analysis tool. %

In this paper, we show the application of the {\tt pygwb} package to mock datasets, illustrating how the various modules can be assembled to form a search pipeline, and showing what a \gls{gwb} detection could look like with our analysis approach. %
To conclude, we run the {\tt pygwb} pipeline on real \gls{gw} data from the third observing run (O3) of the \gls{lvk} collaboration, and recover results in agreement with published results. %
Both analyses serve as a validation of the software.

The {\tt pygwb} package is designed to evolve along the way and address new analysis needs as they arise. %
This is facilitated by the structure and the format of {\tt pygwb}, and the management of the online Git repository. %
The {\tt pygwb} team invites input from the broader community, under the form of Git issues and pull requests. %
New contributors to the code are always welcome. %
Official updates and releases of the code will be handled and reviewed internally by the software and review teams, which are due to evolve. %

We are aware other analysis methodologies exist which accommodate different features of specific \glspl{gwb}, such as potential anisotropy~\cite{Ain:2018zvo}, and the intermittency of the \gls{bbh} background~\cite{Smith:2017vfk, Lawrence:2023buo}. %
We look forward to interfacing with these methods and, where useful and appropriate, improving the current codebase to support and encompass more analysis schemes.

Finally, we are particularly excited at the prospect of broadening the scope of the package to include support for next generation detectors such as \gls{et}~\cite{ET_science} and \gls{ce}~\cite{CE_science}. %
While the science cases and design properties of these detectors are still under development, there is evident interest in targeting \glspl{gwb} with these detectors within the community~\cite{ET-first-MDC-PhysRevD.86.122001, ET-second-MDC-PhysRevD.89.084046, Sathyaprakash:2011bh}, and a notable increase in sensitivity compared to present-day interferometers is expected.  

\section*{Acknowledgements}

The author list of this paper includes, in order: all {\tt pygwb} code authors, in order of successful GitLab merge requests, at time of writing; %
code reviewers and testers, in alphabetical order. %

We would like to thank the LVK stochastic group for its continued support. %
Special thanks to G. Cella and J. Suresh for valuable comments on the manuscript. 

AIR is supported by the NSF award 1912594.
ARR is supported in part by the Strategic Research Program “High-Energy Physics” of the Research Council of the Vrije Universiteit Brussel and by the iBOF “Unlocking the Dark Universe with Gravitational Wave Observations: from Quantum Optics to Quantum Gravity” of the Vlaamse Interuniversitaire Raad and by the FWO IRI grant I002123N “Essential Technologies for the Einstein Telescope”. %
KT is supported by FWO-Vlaanderen through grant number 1179522N. %
PMM is supported by the NANOGrav Physics
Frontiers Center, National Science Foundation (NSF), award number 2020265.
LT is supported by the National Science Foundation through OAC-2103662 and PHY-2011865.
KJ is supported by FWO-Vlaanderen via grant number 11C5720N. %
F.D.L. is supported by a FRIA Grant of the Belgian Fund for Research, F.R.S.-FNRS. %
JL was supported by NSF Award PHY-2207270. %
DD is supported by the NSF as a part of the LIGO Laboratory
AM is supported by the European Union’s Horizon 2020 research and
innovation programme under the Marie Skłodowska-Curie grant agreement No. 754510. %
JDR was supported in part by NSF Award PHY-2207270 and start-up funds provided by Texas Tech University. VM was supported in part by the NSF award PHY-2110238.%

This material is based upon work supported by NSF’s LIGO Laboratory 
which is a major facility fully funded by the 
National Science Foundation.
LIGO was constructed by the California Institute of Technology 
and Massachusetts Institute of Technology with funding from 
the National Science Foundation, 
and operates under cooperative agreement PHY-1764464. 
Advanced LIGO was built under award PHY-0823459.
The authors also gratefully acknowledge the support of the Science
and Technology Facilities Council (STFC) of the United Kingdom, the Max-Planck-Society (MPS), and the
State of Niedersachsen/Germany for support of the construction of Advanced LIGO and construction and
operation of the GEO 600 detector. Additional support for Advanced LIGO was provided by the Australian
Research Council. 
The authors gratefully acknowledge the Italian Istituto Nazionale di Fisica Nucleare
(INFN), the French Centre National de la Recherche Scientifique (CNRS) and the Netherlands Organization for Scientific Research (NWO), for the construction and operation of the Virgo detector and the creation
and support of the EGO consortium. The authors also gratefully acknowledge research support from these
agencies as well as by the Council of Scientific and Industrial Research of India, the Department of Science
and Technology, India, the Science \& Engineering Research Board (SERB), India, the Ministry of Human
Resource Development, India, the Spanish Agencia Estatal de Investigaci\'on (AEI) the Spanish Ministerio de Ciencia e Innovaci\'on and Ministerio de Universidades, the Conselleria de Fons Europeus, Universitat i Cultura and the Direcci\'o General de Pol\'itica Universitaria i Recerca del Govern de les Illes Balears, the Conselleria d’Innovaci\'on, Universitats, Ci\'encia i Societat Digital de la Generalitat Valenciana and the CERCA Programme Generalitat de Catalunya, Spain, the National Science Centre of Poland and the European Union – European Regional Development Fund; Foundation for Polish Science (FNP), the Swiss
National Science Foundation (SNSF), the Russian Foundation for Basic Research, the Russian Science
Foundation, the European Commission, the European Social Funds (ESF), the European Regional Development Funds (ERDF), the Royal Society, the Scottish Funding Council, the Scottish Universities Physics
Alliance, the Hungarian Scientific Research Fund (OTKA), the French Lyon Institute of Origins (LIO), the
Belgian Fonds de la Recherche Scientifique (FRS-FNRS), Actions de Recherche Concertées (ARC) and
Fonds Wetenschappelijk Onderzoek – Vlaanderen (FWO), Belgium, the Paris ˆIle-de-France Region, the
National Research, Development and Innovation Office Hungary (NKFIH), the National Research Foundation of Korea, the Natural Science and Engineering Research Council Canada, Canadian Foundation for
Innovation (CFI), the Brazilian Ministry of Science, Technology, and Innovations, the International Center
for Theoretical Physics South American Institute for Fundamental Research (ICTP-SAIFR), the Research
Grants Council of Hong Kong, the National Natural Science Foundation of China (NSFC), the Leverhulme
Trust, the Research Corporation, the Ministry of Science and Technology (MOST), Taiwan, the United States
Department of Energy, and the Kavli Foundation. The authors gratefully acknowledge the support of the
NSF, STFC, INFN and CNRS for provision of computational resources
The authors are grateful for computational resources provided by the LIGO Laboratory and supported by NSF Grants PHY-0757058 and PHY-0823459. 
This work carries LIGO document number P2300048. %

\appendix
\section{Window functions and bias factors}\label{sec:app_window}
The window factors, $\bar{w}^4_\mathrm{ovl}$ and $\bar{w}^4$, used in Sec.~\ref{sec:postproc} are defined as in Eqs.~(34) and (24) in~\cite{lazz_romano_windowing_note}. They are used to correct for the effect windowing has on our estimate of the variances. Actually, these corrections should include contributions from the autocorrelation function (PSD) of the individual detectors or their cross-correlation (see, e.g. Eqs.~(22) and (32) of the same note). However, if the frequency response of the window is sufficiently strongly peaked around zero, then we can treat the transformed windows as delta functions~\cite{whelan_CC_dcc} and our expressions for these quantities reduce to
\begin{align}
    \bar{w}^4 = \frac{1}{N}\sum_{i=1}^Nw_i^4,
\end{align}
where $w_i$ represents the $i^{\textrm{th}}$ sample of the Hann window we use. Likewise, we need to account for the covariance between point estimates calculated in adjacent time segments. The point estimates are each quadratic in the data, windowed, and use 50\% overlapping segments of data, and so we must account for the overlapping of the windows applied to the two segments
\begin{align}
   \bar{w}_{\textrm{ovl}}^4 = \frac{1}{N/2}\sum_{i=N/2+1}^N w_i^2 w_{i-N/2}^2,
\end{align}
where we see this now as the cross-correlation of the pieces of the two windows that overlap for the two segments.

When calculating the variance of our point estimate, we must estimate the quantity 
$\left(P_{1, f} P_{2, f}\right)^{-1}$, which is the expression that appears in the Gaussian likelihood used to construct our optimal estimators~\cite{Matas:2020roi}, and is therefore the relevant quantity when considering the variance of the point estimates. We briefly summarize how to properly estimate this quantity based on the discussion in Appendix B of ~\cite{Matas:2020roi}, noting that they do not consider the effect of windowing, which we also discuss below. 

For a segment of length $T$ we calculate estimators for the PSDs, $\hat P_{I,f}$, where $I=1,2$ labels the detector, using Welch's method~\cite{welch_method_and_window_factor}. We break our time segment $T$ into 50\% overlapping chunks, calculate the PSD in each chunk, and average those estimates together. If we want a PSD with frequency resolution $\Delta f$ then we have $K$ overlapping segments where $K=2T\Delta f -1$. We can assume our (noisy) estimators for the individual PSDs are unbiased and can be written as the true PSD plus some small deviation, $\hat P_{1,f} = P_{1,f} + \delta P_{1,f}$. We now look at the quantity of interest in calculating our variance
\begin{align}
   \frac{1}{\hat P_{1,f}\hat P_{2,f}} =&\frac{1}{\left[P_{1,f} + \delta P_{1,f}\right]\left[P_{2,f} + \delta P_{2,f}\right]}.
   \end{align}
   We can expand the denominator, take the expectation value of both sides, and use the fact that $\langle\delta P_{I,f}\rangle=0$ and $\langle\delta P_{I,f}^2\rangle = \textrm{var} P_{I,f}$, where $I=1,2$ labels the detector. This gives us
   \begin{align}
    \Braket{\frac{1}{\hat P_{1, f}\hat P_{2, f}}}\approx&\frac{1}{P_{1, f}  P_{2, f}} \left(1 + \frac{\textrm{var}P_{1, f}}{ P_{1, f}^2} + \frac{\textrm{var}P_{2, f}}{ P_{2, f}^2} + \cdots\right)\\
    =&\frac{1}{ P_{1, f}  P_{2, f}}\left(
    1 + \frac{2\kappa}{K} \right).
\end{align}
This expression can be compared to Eq. (B8) in~\cite{Matas:2020roi}, noting that we have an extra term in the variance of our PSDs, $\kappa$. This term reduces the ``effective'' number of averages we perform due to our windowing, where we apply a Hann window with amplitude $\{w_i\}$ at each sample $i$, as well as the overlapping of our chunks of data. The correction factor is given by~\cite{welch_method_and_window_factor}
\begin{align}
   \kappa = \left[1 + 2\left(\frac{\sum_{i=N/2+1}^N w_i w_{i-N/2}}{\sum_{i=1}^Nw_i^2}\right)^2\frac{K-1}{K}\right].
\end{align}
In practice, we ignore the term $(K-1)/K$, as it leads to extra corrections that are $\mathcal {O}(K^{-2})$ that are quite small.

We can now define a bias correction factor based on the windowing we choose and the number of averages used in constructing $\hat P_{I,f}$. 
Defining $N_{\textrm{eff}} = \kappa^{-1}K$, we have
\begin{align}
\hat\sigma^{-2}(f) = \left(1 + \frac{2}{N_\textrm{eff}}\right) \sigma^{-2}(f),
\end{align}
where we have used simplified notation again where the hat indicates our estimator for Eq.~(\ref{eq:Variance}) and the unhatted indicates the true value.

Taking the square root of both sides and inverting it gives us
\begin{align}
    \sigma = b(N_\textrm{eff})\hat\sigma,
\end{align}
where the bias factor, $b(N_{\textrm{eff}})$, is given by
\begin{align}
    b(N_{\textrm{eff}}) = \frac{N_{\textrm{eff}}}{N_{\textrm{eff}}-1},
\end{align}
assuming $N_{\textrm{eff}}$ is large. %
In Sec.~\ref{Sec:DeltaSigma}, two different bias factors are discussed. In one case, the ``naive'' $\sigma$ is estimated using one segment of length $T$, which results in fewer effective averages, and a larger bias correction than our typical estimate of $\sigma$ which uses two adjacent segments of length $T$ and there twice as many averages.

\section{Marginalizing over calibration uncertainty}\label{sec:app_calibration}
Given measurements $\{\hat \Omega_i\}$ with uncertainties $\sigma^2_i$, as shown in Sec.\ref{sec:pe} the following likelihood function can be used to perform parameter estimation on the \gls{gwb}:
	\begin{equation}
    \label{eq:likelihood-again}
    p(\{\hat \Omega_f\} | {\bm \Theta})
    	= \mathcal{N} \exp\left[
        	-\frac{1}{2}\sum_f\frac{\left(\hat \Omega_f -  \Omega_{\rm M}(f|{\bm \Theta})\right)^2}{\sigma_f^2}\right].
    \end{equation}
Here, the $\{\hat \Omega_f\}$ are a set of estimators for the \gls{gw} energy density at discrete frequencies $f$, $\Omega_{\rm M}(f|{\bm \Theta})$ is a model for the energy density with parameters ${\bm \Theta}$, and $\mathcal{N}$ is a normalization constant.
We will consider only a single baseline and neglect the sum over detector pairs $IJ$ appearing in Eq.~\eqref{eq:likelihood}; if multiple detector pairs exist, the derivation below can be replicated for each pair.

Eq.~\eqref{eq:likelihood-again} assumes that our estimators $\{\hat \Omega_f\}$ are direct, unbiased measurements of the underlying energy-density spectrum.
In general, however, the imperfect amplitude and phase calibration of \gls{gw} detectors will break this assumption.
We can account for calibration uncertainty by amending our likelihood to introduce a new parameter $\lambda$:
	\begin{equation}
    \label{eq:likelihood-calibration-uncertainty}
    p(\{\hat \Omega_f\} | {\bm \Theta},\lambda)
    	= \mathcal{N} \exp\left[
        	-\frac{1}{2}\sum_f\frac{\left(\hat \Omega_f -  \lambda\Omega_{\rm M}(f|{\bm \Theta})\right)^2}{\sigma_f^2}\right].
    \end{equation}
The parameter $\lambda$ is an unknown multiplicative factor that encapsulates potential calibration inaccuracy.
In the case of perfect amplitude calibration ($\lambda=1$), then $\{\hat \Omega_f\}$ are direct measurements of the underlying (unknown) energy spectrum.
But if our calibration is imperfect ($\lambda\ne1$), then $\{\hat \Omega_f\}$ are instead measurements of some multiple $\lambda \Omega(f)$ of the \gls{gwb} spectrum.
Although we do not know $\lambda$, it is possible to estimate the \textit{uncertainty} on instrumental calibration.
We will therefore model $\lambda$ itself as an unknown variable drawn from a normal distribution centered at 1 (corresponding to perfect calibration) but with a variance $\epsilon^2$:
	\begin{equation}
    p(\lambda) \propto 
    	\exp\left[-\frac{1}{2\epsilon^2}\left(\lambda-1\right)^2\right],
    \end{equation}
where $\epsilon$ is a known amplitude calibration uncertainty. 
Additionally, we impose the constraint that $\lambda$ be positive: we expect errors in the amplitude of strain measurements but not their \textit{sign}.
In this case, the probability distribution for $\lambda$ becomes
	\begin{equation}
    \label{eq:plambda}
    p(\lambda) = \sqrt{\frac{2}{\pi}}
  		\frac{1}{\epsilon\left[1
        	+\mathrm{Erf}(\frac{1}{\sqrt{2\epsilon^2}})\right]}
        \exp\left[-\frac{1}{2\epsilon^2}\left(\lambda-1\right)^2\right],
    \end{equation}
normalized to unity on the interval $\lambda\in(0,\infty)$.
Eq. \eqref{eq:plambda} is our prior on $\lambda$.

We can now use Eq. \eqref{eq:plambda} to marginalize our likelihood (Eq.~\eqref{eq:likelihood-calibration-uncertainty}) over the unknown calibration factor $\lambda$.
The marginalized likelihood is given by
	\begin{equation}
    \begin{aligned}
    p(\{\hat \Omega_f \} | {\bm \Theta} )
        &= \int p(\{\hat \Omega_f \} | {\bm\Theta},\lambda) \,p(\lambda) d\lambda \\
        &= \mathcal{N} \sqrt{\frac{2}{\pi}}
  			\frac{1}{\epsilon\left[1
        		+\mathrm{Erf}(\frac{1}{\sqrt{2\epsilon^2}})\right]}
            \int_0^\infty \exp\left[
            	-\frac{1}{2}\sum_f \frac{\left(\hat \Omega_f - \lambda\Omega_{\rm M}(f|{\bm \Theta})\right)^2}{\sigma^2_f}
                -\frac{1}{2}\frac{\left(\lambda-1\right)^2}{\epsilon^2}
                \right] d\lambda.
    \end{aligned}
    \end{equation}
If we define 
	\begin{equation}
    A({\bm \Theta}) = \frac{1}{\epsilon^2}+\sum_f\frac{\Omega_{\rm M}(f|{\bm \Theta})^2}{\sigma^2_f},
    \end{equation}
    \begin{equation}
    B({\bm \Theta}) = \frac{1}{\epsilon^2}+\sum_f\frac{\hat \Omega_f \Omega_{\rm M}(f|{\bm\Theta})}{\sigma^2_f},
    \end{equation}
and
	\begin{equation}
    C({\bm \Theta}) = \frac{1}{\epsilon^2}+\sum_f\frac{\hat \Omega^2_f}{\sigma^2_f},
    \end{equation}
the marginal likelihood can be more concisely expressed as
	\begin{equation}
    \label{eq:likelihood-calib-2}
    p(\{\hat \Omega_f \} | {\bm \Theta} )
    	= \mathcal{N} \sqrt{\frac{2}{\pi}}
  			\frac{1}{\epsilon\left[1 +\mathrm{Erf}(\frac{1}{\sqrt{2\epsilon^2}})\right]}
            \int_0^\infty \exp\left[-\frac{1}{2}\left(
            	A({\bm \Theta})\lambda^2 - 2B({\bm \Theta})\lambda + C({\bm \Theta})
                \right)\right] d\lambda;
    \end{equation}
this expression can be analytically integrated to obtain
	\begin{equation}
    p(\{\hat \Omega_f \} | {\bm \Theta} ) =
    	\mathcal{N} \frac{1}{\epsilon\sqrt{A({\bm \Theta})}}
        \left[\frac{1+\mathrm{Erf}(\frac{B({\bm \Theta})}{\sqrt{2A({\bm \Theta})}})}
        	{1+\mathrm{Erf}(\frac{1}{\sqrt{2\epsilon^2}})}\right]
        \exp\left[-\frac{1}{2}\left(C({\bm \Theta})-\frac{B({\bm \Theta})^2}{A({\bm \Theta})}\right)\right].
    \end{equation}

Marginalization of calibration uncertainty is built into the {\tt pygwb\_pe} module, and this calculation is automatically triggered when passing a calibration error $\epsilon\neq 0$. Additional information on the treatment of calibration uncertainties can be found in \cite{Whelan:2012ur}.
\bibliographystyle{aasjournal}
\bibliography{pygwb_bib}

\end{document}